\documentclass[prx,twocolumn,superscriptaddress,notitlepage,nobibnotes,nofootinbib]{revtex4-1}
\usepackage[colorlinks=true, citecolor=magenta, linkcolor=blue, urlcolor=blue]{hyperref}
\usepackage{graphicx}
\usepackage{amsmath}
\usepackage{amssymb}
\usepackage{changes}
\usepackage{amsfonts,bm}
\usepackage{xcolor}
\usepackage{tikz}
\usetikzlibrary{decorations.pathmorphing}
\usetikzlibrary{decorations.markings}
\usepackage{latexsym}   
\usepackage{bbm} 
\usepackage{mathtools}  
\usepackage{placeins}

\allowdisplaybreaks

\newcommand{\bS}{{\bf S}}
\newcommand{\br}{{\bf r}}

\newcommand{\bq}{{\bf q}}
\newcommand{\bx}{{\bf x}}

\makeatletter
\def\maketitle{
\@author@finish
\title@column\titleblock@produce
\suppressfloats[t]
\let\@authors\@empty
\let\and\relax
\let\affiliation\@gobble@opt@one
\let\author\@gobble
\@author@init
\let\@authors\@empty
\let\@authors@curr\@empty
\let\@affil@list\@empty
\let\keywords\@gobble
\let\@keywords\@empty
\let\email\@gobble
\let\@address\@empty
\let\thanks\@gobble
\titlepage@sw{ %
\clearpage
}{}%
}
\makeatother

\begin{document}

\title{Non-linear phononics by equilibrium electric field fluctuations of terahertz cavities}

\author{Emil \surname{Vi\~nas Bostr\"om}}
\thanks{These authors contributed equally to this work \\ \href{mailto:emil.bostrom@mpsd.mpg.de}{emil.bostrom@mpsd.mpg.de}}
\affiliation{Max Planck Institute for the Structure and Dynamics of Matter, Luruper Chaussee 149, 22761 Hamburg, Germany}
\affiliation{Nano-Bio Spectroscopy Group, Departamento de F\'isica de Materiales, Universidad del Pa\'is Vasco, 20018 San Sebastian, Spain}
\author{Marios H. Michael}
\thanks{These authors contributed equally to this work\\ \href{mailto:marios.michael@mpsd.mpg.de}{marios.michael@mpsd.mpg.de}}
\affiliation{Max Planck Institute for the Structure and Dynamics of Matter, Luruper Chaussee 149, 22761 Hamburg, Germany}
\author{Christian Eckhardt}
\affiliation{Max Planck Institute for the Structure and Dynamics of Matter, Luruper Chaussee 149, 22761 Hamburg, Germany}
\author{Angel Rubio}
\email{angel.rubio@mpsd.mpg.de}
\affiliation{Max Planck Institute for the Structure and Dynamics of Matter, Luruper Chaussee 149, 22761 Hamburg, Germany}
\affiliation{Initiative for Computational Catalysis, Flatiron Institute, Simons Foundation, New York City, NY 10010, USA}
\affiliation{Nano-Bio Spectroscopy Group, Departamento de F\'isica de Materiales, Universidad del Pa\'is Vasco, 20018 San Sebastian, Spain}
\date{\today}

\begin{abstract}
Selective excitation of vibrational modes using strong laser pulses has emerged as a powerful material engineering paradigm. However, to have deterministic control over material properties for device applications, an analogous scheme without drive, operating in thermal equilibrium, is desirable. We here propose such an equilibrium analog of the light-driven paradigm, leveraging the strong coupling between lattice degrees of freedom and the quantum fluctuations of the electric field of a THz micro-cavity. We demonstrate this approach by showing, using \textit{ab initio} data, how electric field fluctuations can induce a sub-dominant ferromagnetic order, on top of the dominant zig-zag antiferromagnet order, in FePS$_3$ close to its N\'eel temperature.
\end{abstract}

\maketitle


Dramatic changes to a material's electronic and magnetic orders can be achieved by driving selected phonon modes far out of equilibrium. This scheme, known as non-linear phononics~\cite{Mankowsky2016,Disa2021}, exploits the non-linear coupling between various phonon modes to resonantly drive and selectively break symmetries of the equilibrium system, to favor a non-equilibrium state with qualitatively different properties. For phenomena where the underlying physics is driven by phonon fluctuations, such as ferroelectricity and superconductivity, this can lead to large modifications in the transition temperature~\cite{Nova2019,Xian19,Budden2021,Disa2021,Rowe2023,Eckhardt24,Sam24}. Similarly, magnetic orders in magnetic insulators are sensitive to small changes in ionic positions~\cite{Zhang2021,Cui2023,Truc2023,Disa2023,Ergeen2023}, because spin interactions typically arise from an interplay of local correlations and virtual kinetic processes, which depend strongly on the bond angles and bond lengths of the crystal~\cite{Jackeli2009}.

Although non-linear phononics has been established as a powerful paradigm to manipulate matter, a general drawback of the scheme is that it operates within a non-equilibrium setting. The effect is therefore short lived, and typically lasts on the order of $10 - 100$ ps. Although in some cases the effects induced by non-linearly driven phonons have been found to be quasi-stationary~\cite{Truc2023,Disa2023,Ilyas2023}, with a lifetime extending orders of magnitude beyond the microscopic timescales, the true equilibrium state remains unchanged. To fully unlock the engineering potential of controlling material structures with light, non-linear phononics must be extended to thermal equilibrium.

In this Letter, we make progress in this direction by using that the non-linear mechanism of shifting the position of a Raman active phonon through laser pumping is an incoherent process~\cite{Foerst11}. As a result, the same result can be achieved through electric field noise of the same magnitude. We show that a realistic route towards replacing intense laser light is to use thermal or quantum electric field fluctuations of THz cavities, and provide a concrete example for the antiferromagnet FePS$_3$.

Our work is part of a larger effort to use optical micro-cavities to effect material control in thermal equilibrium~\cite{Latini2021,Jon23,Hector24,Appugliese2022,VinasBostrom2023,Dirnberger2023,Jarc2023,Riccardo24,Chris_arXiv,Bloch2022}, resting on the idea that changing a materials dielectric environment can alter its properties~\cite{Hubener2024}. Recent experiments have demonstrated this effect and shown that it is highly off-resonant, and therefore sensitive to changes in the electromagnetic mode structure at all frequencies~\cite{Appugliese2022,Jarc2023}. This type of ground state reconfiguration is therefore qualitatively different from polaritonic cavity physics, where selected cavity modes are driven and resonantly coupled to excitations of matter~\cite{Flick2017,Sentef2018,Thomas2019,Dirnberger2022,Bae2022,Dirnberger2023,KamperSvendsen2023}. To capture such off-resonant effects it is necessary to develop a theoretical framework that accounts for the full mode structure of the electromagnetic field, as well as their thermal population.

\begin{figure}
 \centering
 \begin{tikzpicture}

  \begin{scope}[shift={(0.2,0.5)},rotate=0,scale=1.0]
   \shade[left color=lightgray, right color=lightgray, middle color=white, shading angle=90] (0,-0.8) rectangle (4,-0.6);
   \shade[top color=white, bottom color=red, middle color=white, shading angle=0, opacity=0.2] (0.3,-0.6) -- (3.7,-0.6) to [bend left=10] (3.7,0.6) -- (0.3,0.6) to [bend left=10] (0.3,-0.6);
  \end{scope}

  \draw[<->, >=stealth] (0.4,-0.02) -- (0.4, 0.40);
  \node at (0.2, 0.24) {\footnotesize $d$};

  \node at (-2.3, 0.0) {\includegraphics[width=0.4\columnwidth]{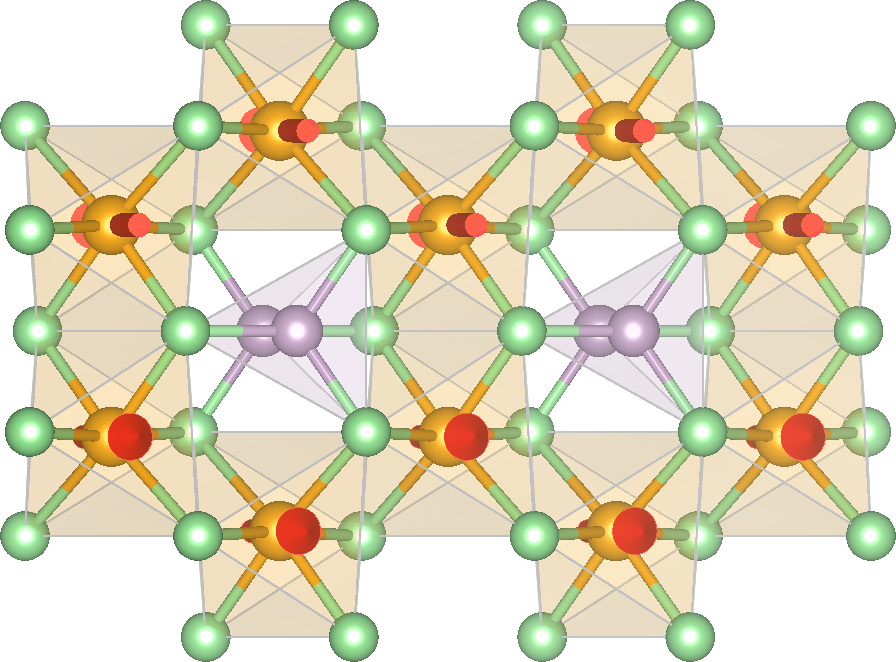}};
  \node at ( 2.15,0.9) {\includegraphics[width=0.4\columnwidth]{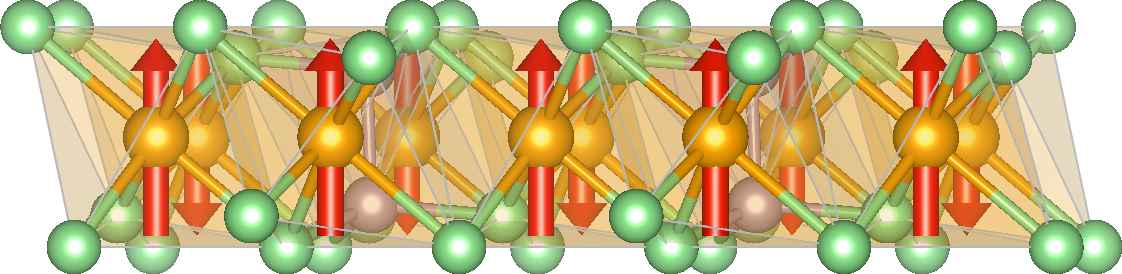}};

  \begin{scope}[shift={(-2.1,-1.19)},rotate=0,scale=1.0]
   \draw[dashed] (0.0, 0.0) -- (1.3, 0.0) -- (1.3, 2.38) -- (0.0, 2.38) -- (0.0, 0.0);
  \end{scope}

  \node at (0.5,-0.7) {\footnotesize $\langle E^2\rangle > 0$};
  \node at (2.3,-0.7) {\footnotesize $Q > 0$};
  \node at (3.8,-0.7) {\footnotesize $M > 0$};

  \draw[decoration={markings, mark=at position 0.6 with {\arrow{stealth}}},postaction={decorate}] (-0.5, -1.35) -- (0.08, -1.35);
  \draw[decorate,decoration={snake,amplitude=1.0,segment length=5}] (0.5,-1.4) circle (0.4);

  \draw[black,dashed] plot[smooth,domain=1.8:2.6] (\x, {4*(\x-2.2)*(\x-2.2)-1.65});
  \draw[black] plot[smooth,domain=2.0:2.8] (\x, {4*(\x-2.4)*(\x-2.4)-1.75});

  \draw[black,dashed] plot[smooth,domain=3.3:4.1] (\x, {4*(\x-3.7)*(\x-3.7)-1.65});
  \draw[black] plot[smooth,domain=3.5:4.3] (\x, {4*(\x-3.9)*(\x-3.9)-1.75});

  \draw[->,>=stealth,color=red] (1.1,-1.55) -- (1.8,-1.55);
  \node at (1.45,-1.8) {$\lambda$};
  \draw[->,>=stealth,color=red] (2.8,-1.55) -- (3.3,-1.55);
  \node at (3.1,-1.8) {$g$};

  \node at (-4.1, 1.2) {\bf a};
  \node at ( 0.2, 1.0) {\bf b};
  \node at (-0.28,-1.05) {\bf c};
 \end{tikzpicture}
    \caption{({\bf a}) Top view of a monolayer of the van der Waals antiferromagnet FePS$_3$, showing the Fe (orange), P (purple) and S (green) ions, as well as the zig-zag antiferromagnetic order of the magnetic moments (red). The black dashed lines indicate the magnetic unit cell. ({\bf b}) Illustration of FePS$_3$ deposited on a plasmonic substrate, with the substrate-material distance $d$ indicated. For a plasmonic cavity, the effective mode volume is related to the substrate-material distance by $V_{\rm eff} = d^3$. ({\bf c}) Illustration of how electric field fluctuations generate a net phonon displacement, which in turn generates a net magnetization.}
    \label{fig:schematic}
\end{figure}

We here propose a scheme to control the properties of two-dimensional quantum materials through their coupling to the fluctuations of a confined electromagnetic field, which we term equilibrium non-linear phononics. This field can arise either from surface phonon polaritons (SPPs) close to a paraelectric insulating substrate~\cite{Eckhardt24}, surface plasmon polaritons~\cite{Lenk22,kipp2024} close to a plasmonic substrate, or from an optical micro-cavity. In contrast to standard non-linear phononics, this scheme operates in global thermal equilibrium. More specifically, by coupling a Raman-active phonon mode to the vacuum and thermal fluctuations of a confined electric field, we can effect deterministic control over the phonon displacement. To demonstrate the power of this approach, we show how a sub-dominant ferromagnetic magnetization appears in the van der Waals antiferromagnet FePS$_3$, when tuned close to its N\'eel temperature, due to the collective coupling between spins and phonons, and phonons and SPPs. This magnetization is due to a cavity-mediated asymmetry of the free energy landscape of FePS$_3$, which is amplified by critical magnetic fluctuations appearing close to the antiferromagnetic transition~\cite{Ilyas2023}.


The non-linear phononics scheme proceeds by identifying a material property $M$ that is sensitive to the displacement of a phonon mode $Q$, such that $M = 0$ for $Q = 0$ but acquires a finite value for $Q \neq 0$. A net displacement of $Q$ is achieved through a non-linear coupling to some other (typically infra-red active) phonon mode $Q_{\rm IR}$, of the general form $g Q Q_{\rm IR}^2$. When the IR mode is driven by an external electric field, $Q_{\rm IR}$ displays an oscillatory motion that persists during the phonon lifetime $\tau$. The oscillatory motion of $Q_{\rm IR}$ generates a net force on the coordinate $Q$, leading to a net displacement whose magnitude and direction is set by the coupling constant $g$. Unless $M$ or $Q$ is trapped in a new meta-stable state, the effect will decay after a time $t \sim \tau$, typically on the order of $10 - 100$ ps.

The key feature of this scheme is the net force on $Q$ effected via the IR-active phonon $Q_{\rm IR}$. A similar force can be generated through the quadratic coupling to an electric field, $\lambda QE^2$, which however only persists during the action of the pulse. The situation changes drastically when the electric field of the laser is replaced by the fluctuating electric field of a confined electromagnetic mode. We can therefore imagine replacing the Hamiltonian of a typical non-linear phononics setup
\begin{align}
 H &= H_{\rm e-p} + \sum_i g_i Q Q_{{\rm IR}, i}^2 + \sum_i Z_{ij} Q_{{\rm IR}, i} E_j,
\end{align}
where $H_{\rm e-p}$ is the electron-phonon Hamiltonian, $g_i$ is the non-linear phonon-phonon coupling, and $Z_{ij}$ is the Born effective charge tensor for the IR-mode $i$, with the Hamiltonian
\begin{align}
 H &= H_{\rm e-p} + \sum_i \lambda_i Q \hat{E}_i^2 + \sum_i \hbar\omega_i a_i^\dagger a_i.
\end{align}
Here $\hat{E}_i$ is the electric field of a quantized electromagnetic mode $i$, with frequency $\omega_i$, whose corresponding creation and annihilation operators are $a_i^\dagger$ and $a_i$. The parameter $\lambda_i$ is the quadratic light-matter coupling, related to the Raman tensor of mode $Q$, and quantifies the interaction between the phonon mode and the electric field fluctuations.

For a given phonon mode $Q$, the quantized electric field effects a displacive force $\sum_i \lambda_i \langle \hat{E}_i^2\rangle$ stretching the mode away from its equilibrium configuration. This force is proportional to the mode confinement $1/V_{\rm eff}$, which for a micro-cavity is set by the effective mode volume and for SPPs by the cube of the substrate-material distance (see Fig.~\ref{fig:schematic}). The equation above illustrates that if the effective force is appreciable, the fluctuations of the confined electromagnetic mode will have the same effect on the material as the non-linear phonon coupling, but in thermal equilibrium. Such fluctuations are expected to be rather insensitive to the precise cavity geometry, and could be realized e.g. with a distributed Bragg reflector~\cite{Dirnberger2022,Dirnberger2023}, a split-ring resonator~\cite{Appugliese2022}, or other types of cavities~\cite{Hubener2020}. For concreteness however, we focus in the remainder of this work on the electric field coming from SPPs, and show how their coupling to a Raman phonon leads to a finite magnetization in the van der Waals magnet FePS$_3$.


To calculate the fluctuations of the phonon-polariton electric field, we assume a paraelectric substrate hosting a phonon-polariton whose normal to the surface is along the $z$-direction. The components of the electric field operator parallel ($\parallel$) and perpendicular ($\perp$) to the surface can then be expanded in terms of polaritonic mode functions as
\begin{align}
 {\bf E}(\br,t) &= \sum_\bq {\bf f}_\bq e^{-qz} e^{i \bq \cdot \bx - i \omega_s t}  \left( a^\dagger_\bq + a_{-\bq} \right).
\end{align}
Here $\omega_s$ is the SPP frequency, and ${\bf f}_\bq$ is a mode function depending on the transverse optical (TO) and longitudinal optical (LO) phonon frequencies $\omega_{\rm TO}$ and $\omega_{\rm LO}$, as well as on the relative permittivities $\epsilon_{\rm sub}$ and $\epsilon_{\rm mat}$ of the substrate and the material. Since the SPP dispersion is approximately flat as a function of $\bq$ (see Fig.~\ref{fig:plasmon_coupling}a), we have replaced the frequency $\omega_\bq$ with $\omega_s$, and note that due to the longitudinal polarization of the phonon-polariton modes the electric field ${\bf E} \parallel \bq$ for the in-plane components. A detailed discussion of the quantization and orthonormalization of the phonon-polariton modes is found in Ref.~\cite{Supplemental}.

For our present purposes, there are two important benefits with SPPs. The first is the geometry of their confinement, localizing the SPPs to a region of height $d = q^{-1}$ above the paraelectric surface. For the interaction with a 2-D system, this confinement gives a close to optimal overlap between the modes and the material. The second benefit is the dispersion of the modes, with its almost flat $\omega_\bq = \omega_s$ for $|\bq| > q_c$ (see Fig.~\ref{fig:plasmon_coupling}a). This gives a large density of states at $\omega = \omega_s$ (see Fig.~\ref{fig:plasmon_coupling}b), whose effect is to produce large electric field fluctuations.

\begin{figure}
 \centering
 \begin{tikzpicture}
  \node at (-2.25, 2.1) {\includegraphics[width=0.5\columnwidth]{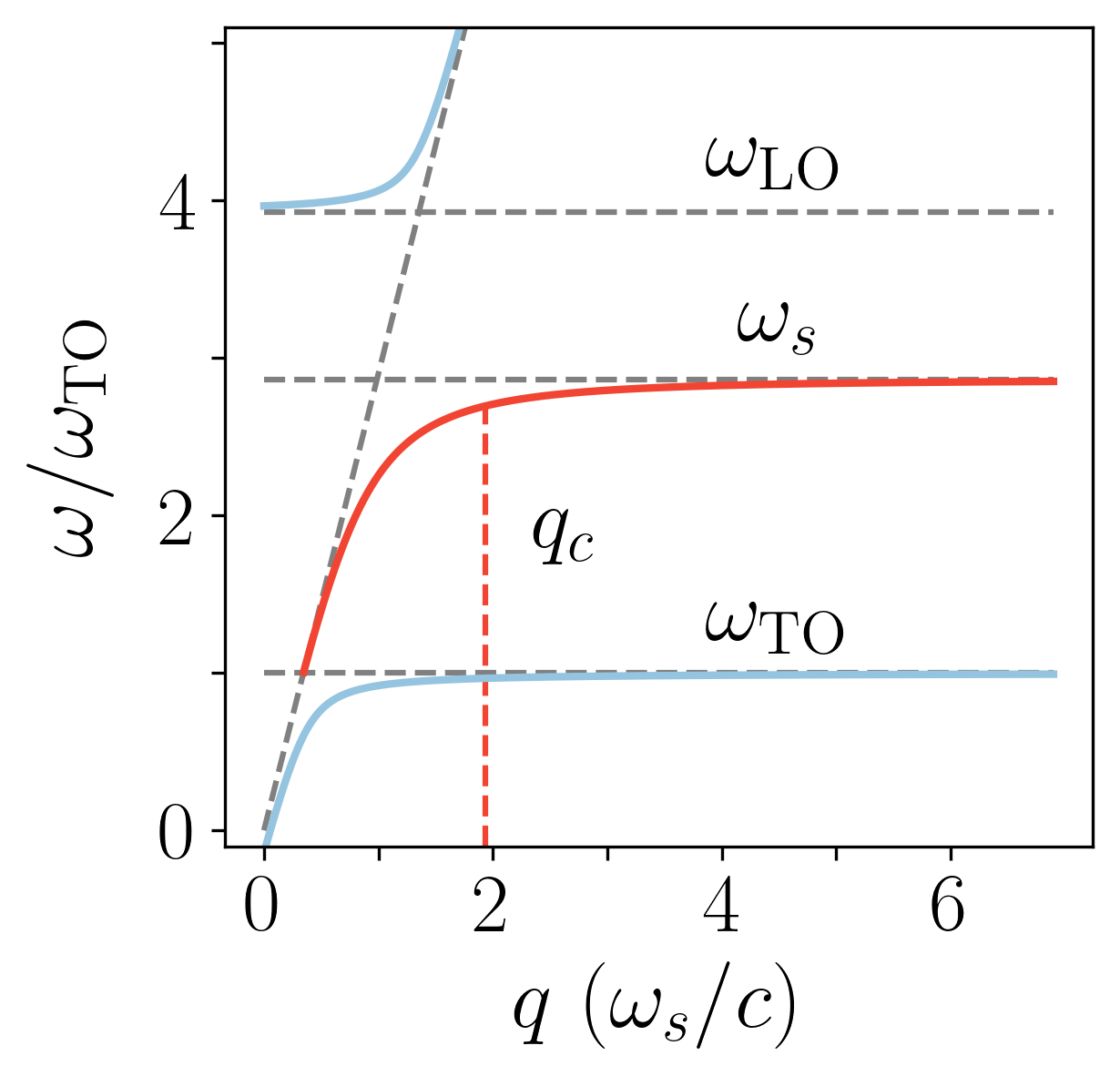}};
  \node at ( 2.15, 2.1) {\includegraphics[width=0.5\columnwidth]{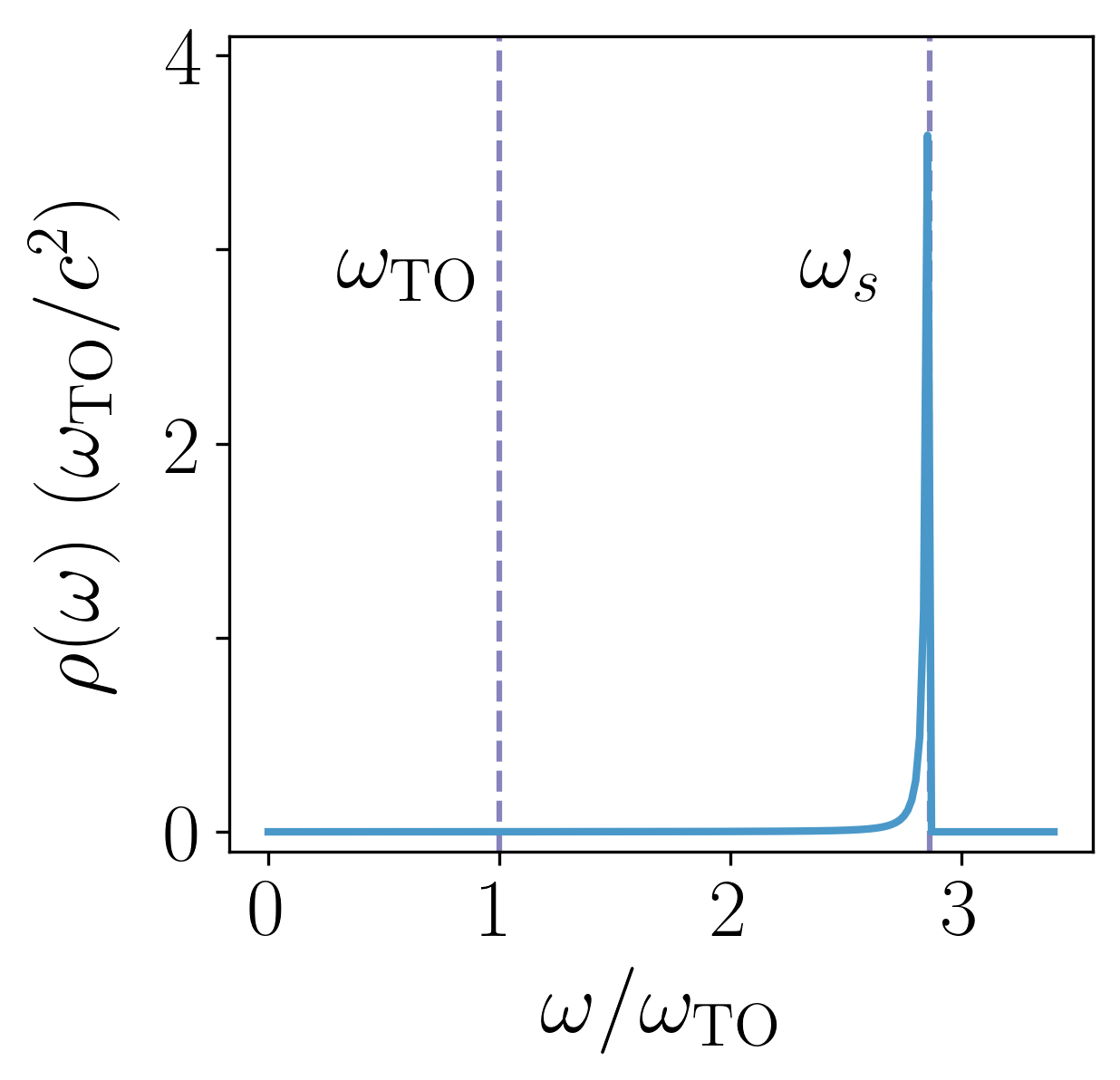}};
  \node at (-2.25,-2.0) {\includegraphics[width=0.5\columnwidth]{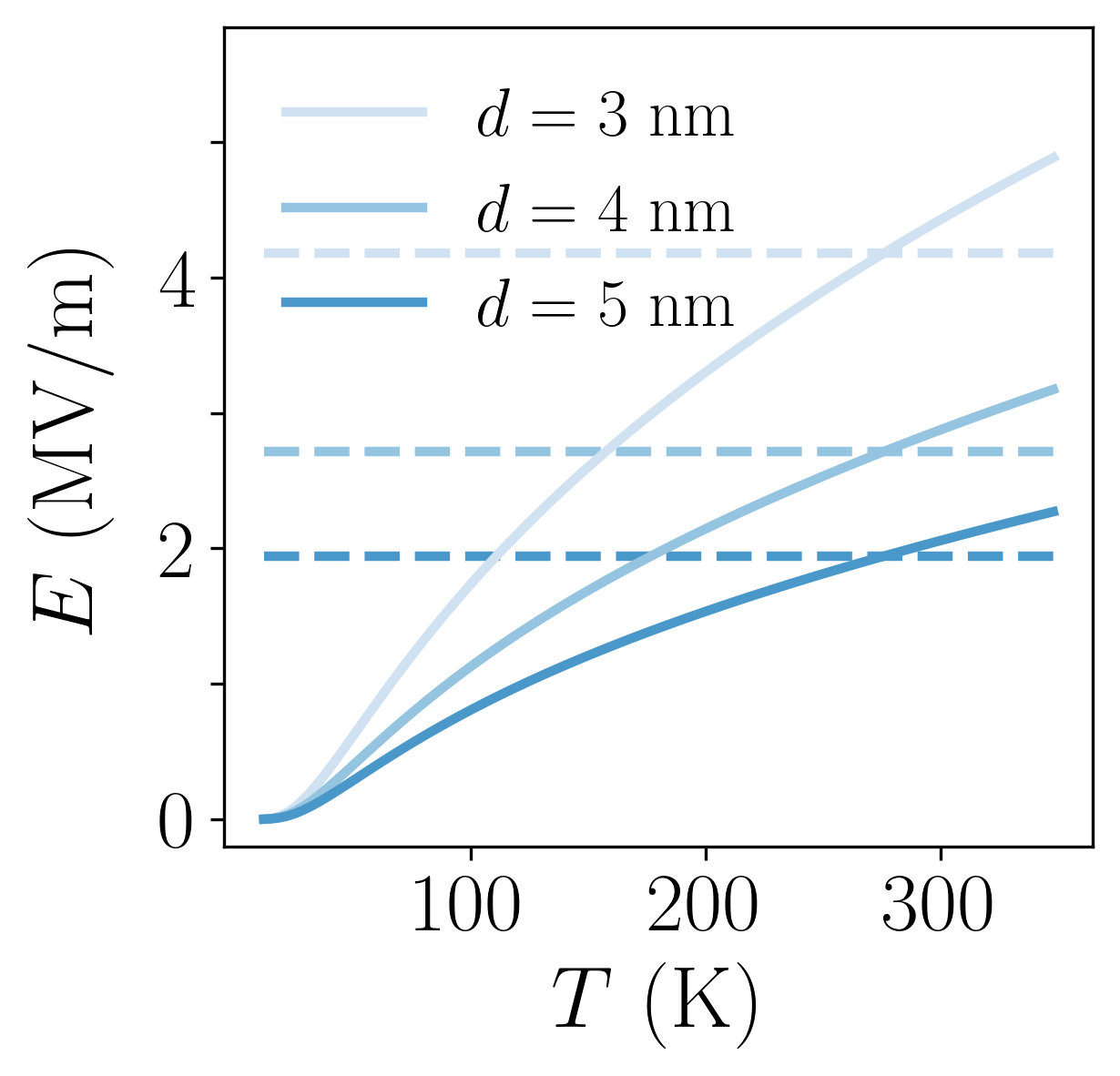}};
  \node at ( 2.15,-2.0) {\includegraphics[width=0.5\columnwidth]{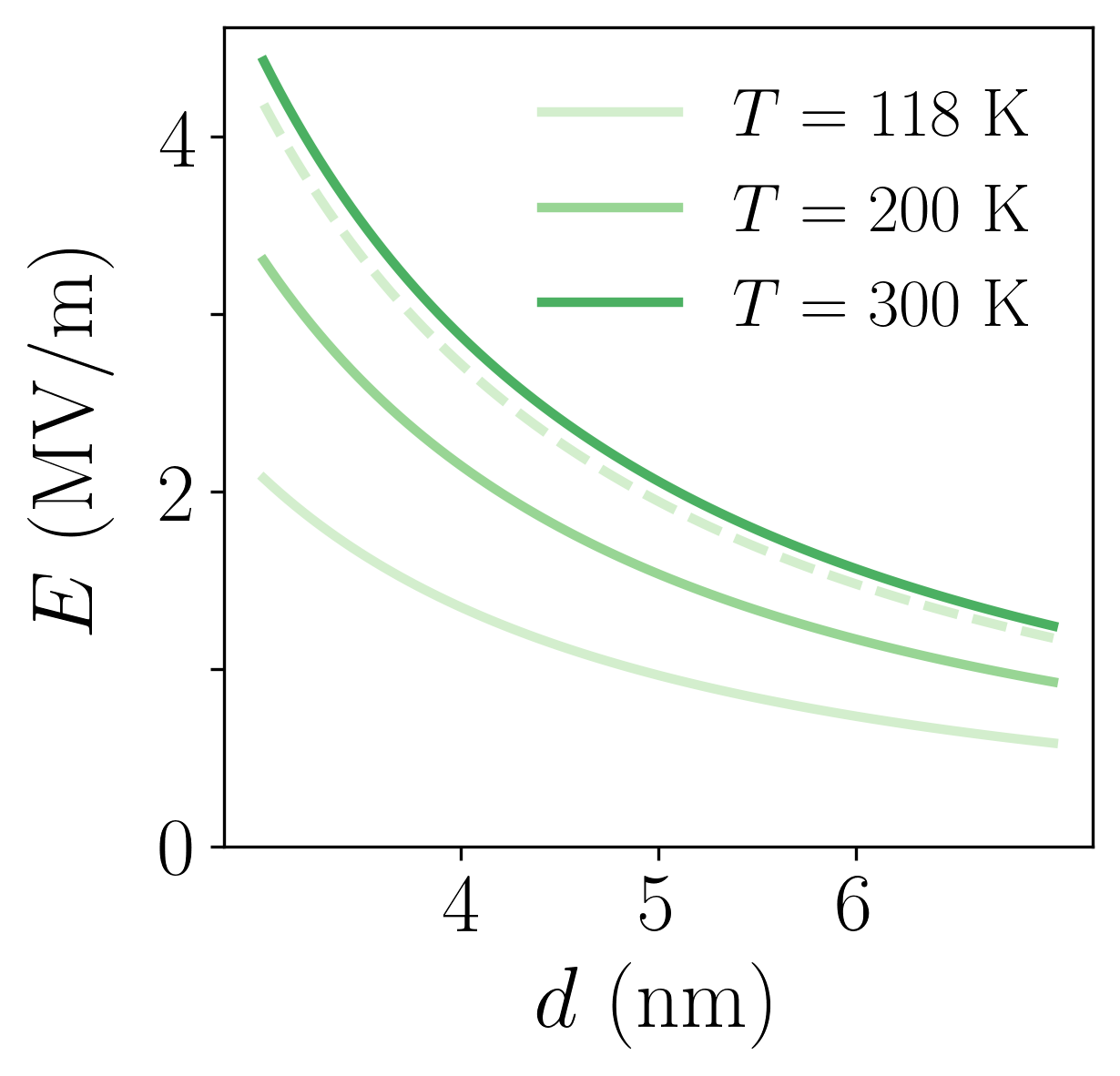}};

  \node at (-3.2, 3.7) {\bf a};
  \node at ( 1.2, 3.7) {\bf b};
  \node at (-0.8,-0.3) {\bf c};
  \node at ( 1.6,-0.4) {\bf d};
 \end{tikzpicture}
 \caption{({\bf a}) Surface phonon polariton (SPP) dispersion (red) as a function of momentum. ({\bf b}) SPP density of states as a function of frequency. ({\bf c}) Temperature dependence of the SPP electric field fluctuations for distances $d = 3$, $4$ and $5$ nm (light blue to dark blue). The dashed lines indicate the vacuum contribution to $E = \sqrt{\langle E^2 \rangle}$, assuming the ultraviolet contribution is unchanged by the paraelectric substrate. ({\bf d}) Distance dependence of the SPP electric field fluctuations for temperatures $T = 118$, $200$ and $300$ K (light green to dark green). The dashed line indicates the vacuum contribution. The electric field $E$ is evaluated for the parameters $\omega_s/2\pi = 2.9$ THz, $\omega_{\rm TO}/2\pi = 1.3$ THz, $\epsilon_{\rm sub} = 6.3$ and $\epsilon_{\rm mat} = 1$.}
 \label{fig:plasmon_coupling}
\end{figure}


At a given temperature $T$, the electric field fluctuations of SPPs stretching a Raman phonon mode are given by
\begin{align}
 \epsilon_0 \sum_{ij} \langle E_i E_j \rangle_{T} &= \frac{\omega_s^2 -\omega_{\rm TO}^2}{16\pi\omega_s (\epsilon_{\rm sub} + \epsilon_{\rm mat}) d^3} n(\omega_s),
\end{align}
where $d$ the is the distance between the substrate and the material and $n(\omega)$ is the Bose-Einstein distribution. The dependence of the fluctuations $E = \sqrt{\langle E^2\rangle}$ on $d$ and $T$ is shown in Fig.~\ref{fig:plasmon_coupling}. We note that this coupling is highly off-resonant, and involves a summation over the complete set of SPP modes contributing to the local fluctuations.

To derive the above expression, we calculated the thermal fluctuations using the expression ${:E^2:} = a^2 + (a^\dag)^2 + 2 a^\dag a = E^2 - \langle E^2_{\rm vac} \rangle$, normal ordered with respect to the phonon-polariton vacuum. It remains a subtle question whether the difference in vacuum fluctuations between free space and a phonon-polariton surface contributes to the stretching of the Raman phonon, as both terms in $\langle E^2_{\rm vac}\rangle - \langle E^2_{{\rm vac},0}\rangle$ formally diverge with the ultraviolet cut-off $\Lambda$. While we expect these divergences to cancel, yielding a finite positive contribution, the challenge of regularizing the theory in a physically consistent way remains an open question. In Fig.~\ref{fig:plasmon_coupling} we show an estimate of the vacuum contribution, calculated by assuming that the mode spectrum above $\omega \sim \omega_s$ is unchanged by the paraelectric substrate. We note that the effective light-phonon coupling is even larger if quantum fluctuations also contribute to this coupling.

Interestingly, a qualitatively similar result is obtained for the noise of the electric field parallel to the mirrors of an idealized Fabry-Perot cavity~\cite{Supplemental}. Indeed, in this case we find $\epsilon_0 \langle E^2 \rangle = 7\pi^2\omega_c/(1080h^3)$ where $\omega_c$ is the fundamental cavity frequency and $h$ the height of the cavity. For both the SPPs and the Fabry-Perot cavity we thus find that the electric field fluctuations are given by the ratio of a characteristic frequency ($\omega_s$ or $\omega_c$) to a characteristic volume ($d^3$ or $h^3$). Given the disparate nature of these cavities, we believe this to be a general result, such that the fluctuations of a cavity can be estimated from its characteristic frequency and length scales. We note that this is a non-trivial result, since the volume of each independent mode is infinite due to the macroscopic in-plane extent of the cavities. However, when all modes are summed their contribution is finite, highlighting the importance of a multi-mode treatment.

We now discuss the effect of SPP fluctuations on FePS$_3$, whose magneto-elastic properties are described by the Hamiltonian~\cite{Liu2021,Zhang2021}
\begin{align}\label{eq:microscopic_ham}
 H &= \sum_{ij} J_{ij} {\bf S}_i \cdot {\bf S}_j - \Delta \sum_{ij} S_i^z S_j^z + \sum_\alpha \frac{\Omega_\alpha}{2} \Big[ P_\alpha^2 + Q_\alpha^2 \Big].
\end{align}
Here the parameters $J_{ij}$ encode the coupling of spins ${\bf S}_i$ and ${\bf S}_j$, $\Delta$ is an Ising anisotropy, $P_\alpha$ and $Q_\alpha$ is the momentum and coordinate of the phonon mode $\alpha$, and $\Omega_\alpha$ is the corresponding phonon energy. The magnetic interactions depend on the ionic positions, which generates an interaction between the spins and lattice displacements. For small displacements $Q$, the exchange interaction can be expanded as $J_{ij}(Q) \approx J_{ij} - \alpha_{ij} Q$, where $\alpha_{ij}$ quantifies the phonon modulation of the magnetic interactions. All parameters of this Hamiltonian have been calculated from first principles within the frozen phonon approximation~\cite{Ilyas2023,Supplemental}. The elements of $\alpha_{ij}$ were found to decay rapidly with inter-atomic distance, and to be significant only for nearest neighbor spins.


Below its N\'eel temperature $T_{\rm N} = 118$ K, FePS$_3$ orders into a zig-zag antiferromagnet (AFM) as illustrated in Fig.~\ref{fig:schematic}a. It was recently observed that when subjected to intense THz pulses, at temperatures close to $T_{\rm N}$, the system develops a sub-dominant ferromagnetic (FM) order on top of the antiferromagnetic order~\cite{Ilyas2023}. The THz-induced magnetization is linked to the non-linear excitation of a Raman-active phonon mode at $\Omega = 3.27$ THz, which interacts strongly with the magnetic structure.

To describe the interaction between this Raman phonon mode and the magnetic orders, the Hamiltonian in Eq.~\ref{eq:microscopic_ham} can be mapped onto an effective theory in terms of the phonon coordinate $Q$, and the macroscopic magnetic variables $L$ and $M$ corresponding to the AFM and FM orders (related to the microscopic spins as discussed in Refs.~\cite{Ilyas2023,Supplemental}). In the effective theory, the system is described by a Ginzburg-Landau free energy
\begin{align}\label{eq:ginzburg_landau}
 F &= \frac{a_L(T)}{2} L^2 + \frac{b_L}{4} L^4 + \frac{a_M(T)}{2} M^2 + \frac{b_M}{4} M^4 \\
 &+ \frac{\Omega}{2} Q^2 + g (L -\langle L\rangle) M Q. \nonumber
\end{align}
The general form of the coupling follows from symmetry, by noting that both $L$ and $Q$ are finite momentum orders ($Q$ is a zone-folded phonon), and that $L$ and $M$ are both odd under inversion. The specific form of the coupling has been corroborated by spin Monte Carlo simulations of the magnetic structure, with the microscopic magnetic interactions modulated according to the phonon displacement pattern~\cite{Ilyas2023,Supplemental}. The real space motion of the atoms corresponding to this phonon mode modifies the bond lengths between the Fe atoms, which enhances the exchange interaction within every other FM zig-zag chain, while weakening it within adjacent chains (see Fig.~\ref{fig:magnetization}). The effect of the modulation is to decrease or increase the magnetic fluctuations $\Delta S$ in every other chain, leading to a net magnetization from the fluctuating moments. Far above or far below the N\'eel temperature, where there is no magnetic order or it is completely frozen in, the effect of the phonon is negligible.


The phonon mode $Q$ is Raman active, and therefore couples to the fluctuations of the cavity electric field via the Raman tensor. The coupling is of the form~\cite{Supplemental}
\begin{align}
 H_R &\approx \frac{\epsilon_0 V_{\rm c}}{2} \hat{\bf E} \frac{\partial \boldsymbol\epsilon}{\partial Q_i} \hat{\bf E} \, \hat{Q}_i,
\end{align}
where $V_{\rm c}$ is the unit cell volume, $\boldsymbol\epsilon$ is the relative permittivity matrix, and all operators are evaluated in a given unit cell $i$. We note that when transitioning from the local picture to a picture in terms of the macroscopic variable $Q$ (the $q = 0$ component of the phonon mode), the Raman tensor is suppressed by a factor $1/\sqrt{N}$ while $Q_i = \sqrt{N} Q$. The coupling to the macroscopic mode therefore remains finite in the thermodynamic limit. Introducing the phonon operator $\hat{Q} = (l_0/\sqrt{2}) (b^\dagger + b)$, where $l_0 = \sqrt{\hbar/(M\Omega)}$ is the phonon oscillator length and $M$ its effective mass, the coupling can be written as $H_R = \lambda (L/L_s)(b^\dagger + b)$ with the effective Raman coupling $\lambda = (\epsilon_0 V_{\rm c}l_0)/(2\sqrt{2}) \langle \hat{\bf E} \partial_Q\boldsymbol\epsilon \hat{\bf E} \rangle$. Here the factor $L/L_s$, with $L_{\rm s}$ the saturation magnetization, is included to account for the fact that the sign of $Q$ is defined relative to the sign of $L$~\cite{Ilyas2023}.


This expression for $\lambda$ holds irrespective of the origin of the electric field, but for a numerical estimate we evaluate the local electric field fluctuations arising from the SPPs of SrTiO$_3$. For a substrate-material distance $d = 1$ nm and a temperature $T = T_{\rm N} = 118$ K, the electric field fluctuations of the vacuum are found to be approximately $E = \sqrt{\langle E^2\rangle} \approx 21.7$ MV m$^{-1}$, while the full temperature dependent field is $E(T) \approx 30.0$ MV m$^{-1}$. We note that this field strength is comparable to the peak value used in recent experiments~\cite{Ilyas2023}, such that a finite effect on the system is expected. With the Raman tensor $\partial\boldsymbol\epsilon/\partial Q_i = 4.2 a_0^{-1}$, unit cell volume $V_{\rm c} \approx 10^3$ {\AA$^3$} and oscillator length $l_0 = 0.35 a_0$, as obtained from first principles simulations~\cite{Supplemental}, this gives an effective coupling of $\lambda \approx 26$ $\mu$eV stretching the phonon by an amount $\Delta Q = 0.34$ m{\AA}.


Having calculated the phonon-cavity coupling, the Ginzburg-Landau theory can now be used to investigate the effect of the SPPs on the magnetic order. Minimizing the phonon energy gives a displaced equilibrium position $Q_0 = \lambda L/(\Omega L_s) - g(L - \langle L\rangle) M/\Omega$, which will influence the magnetization. To calculate the induced magnetization, the Ginzburg-Landau equation for $M$ is solved to lowest order in $\lambda/\Omega$. Introducing for simplicity the coefficient $c_M(T) = g\lambda (\langle L^2\rangle - \langle L\rangle^2)/(\Omega L_s)$, the location of the free energy minimum is given by $M_0 = c_M/\bar{a}_M$. Here $\bar{a}_M(T) = a_M(T) - (g^2/\Omega) (\langle L^2\rangle - \langle L\rangle^2)$ is a renormalized quadratic term for $M$, arising from the coupling between $M$ and $L$ by the phonon. Since $L$ is the dominant order parameter, the coefficient $a_L(T) = a_{L0} (T - T_\textrm{N})^\beta$ will change sign at the N\'eel temperature, leading to a finite value $\langle L\rangle \sim (T_\textrm{N} - T)^\beta$. Similarly, the fluctuations of $L$ will show a critical enhancement at the N\'eel temperature, such that $\chi_L = V(\langle L^2\rangle - \langle L\rangle^2) \sim |T - T_\textrm{N}|^{-\gamma}$.

\begin{figure}
 \begin{tikzpicture}
    \node at (0.0,-2.3) {\includegraphics[width=1.0\columnwidth]{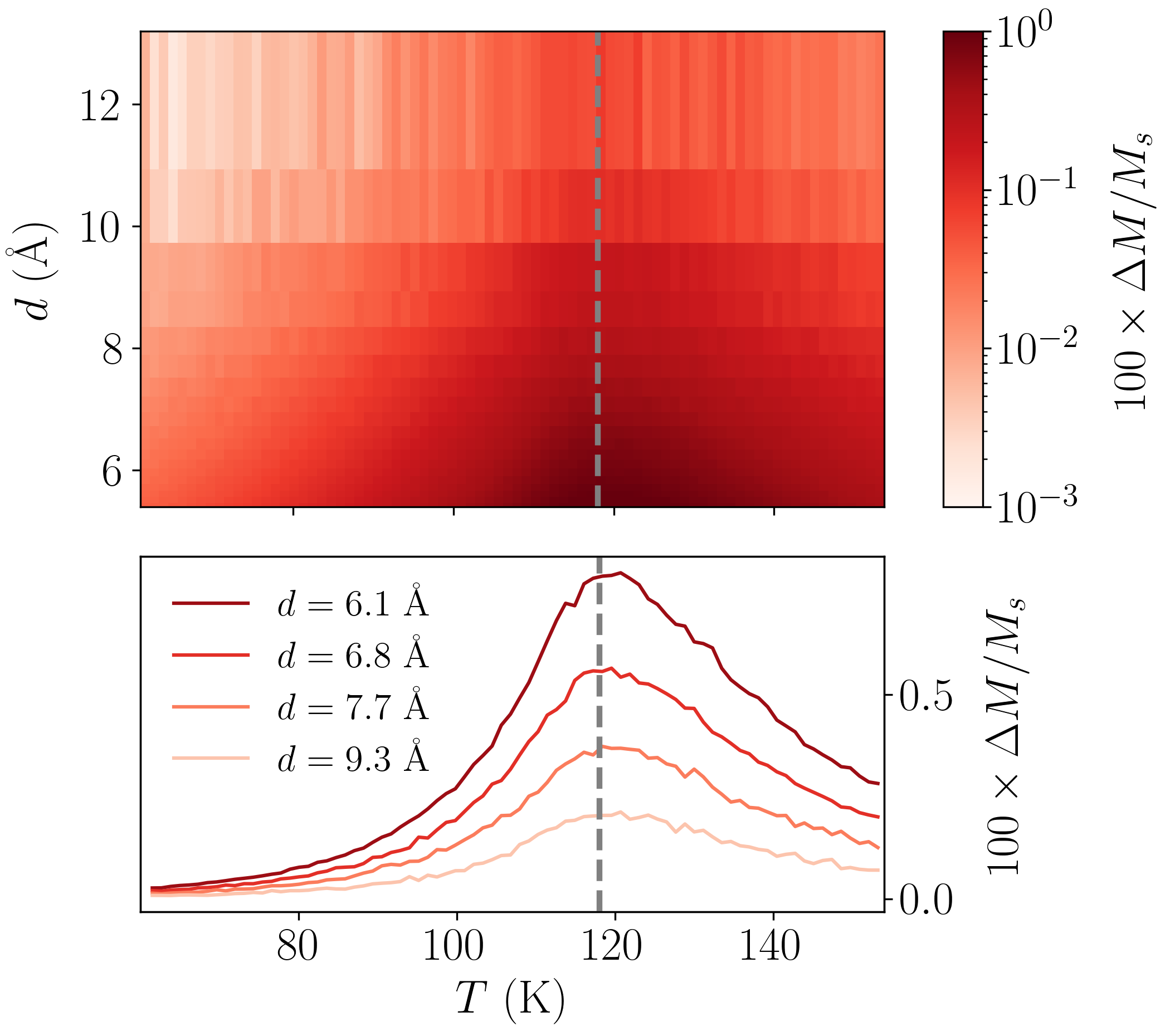}};
     
    \newdimen\R
    \R=0.5cm
            
    \begin{scope}[shift={(-2.4,-0.1)},rotate=0,scale=1.4]
      \draw[line width=0.10mm] (-30:\R) \foreach \x in { 30} { -- (\x:\R)};
      \draw[line width=0.10mm] (150:\R) \foreach \x in {210} { -- (\x:\R)};
      \draw[line width=0.50mm, color=red!60!white] ( 30:\R) \foreach \x in { 90,150} { -- (\x:\R)};
      \draw[line width=0.20mm, color=blue!60!white ] (210:\R) \foreach \x in {270,330} { -- (\x:\R)};
      \foreach \x in {30,90,150}
      \node[inner sep=2.0pt, circle, draw, color=red!80!white, fill=red!80!white] at (\x:\R) {};
      \foreach \x in {210,270,330}
      \node[inner sep=2.0pt, circle, draw, color=blue!80!white , fill=blue!80!white ] at (\x:\R) {};
      
      \draw[->,>=stealth,line width=0.30mm, color=red!80!white, opacity=0.8] (0.0, \R) -- (0.0,1.6*\R);
      \draw[->,>=stealth,line width=0.30mm, color=red!80!white, opacity=0.5] (0.0, \R) -- ( 0.05,1.55*\R);
      \draw[->,>=stealth,line width=0.30mm, color=red!80!white, opacity=0.5] (0.0, \R) -- (-0.05,1.55*\R);

      \draw[->,>=stealth,line width=0.30mm, color=blue!80!white, opacity=0.8] (0.0,-\R) -- (0.0,-1.6*\R);
      \draw[->,>=stealth,line width=0.30mm, color=blue!80!white, opacity=0.5] (0.0,-\R) -- ( 0.05,-1.55*\R);
      \draw[->,>=stealth,line width=0.30mm, color=blue!80!white, opacity=0.5] (0.0,-\R) -- (-0.05,-1.55*\R);

      \node at (0.88,-0.6) {\footnotesize $\langle S\rangle =-S + \Delta S$};
      \node at (0.80, 0.6) {\footnotesize $\langle S\rangle = S - \Delta S$};
    \end{scope}

    \node at ( 1.8, 0.9) {\bf a};
    \node at ( 1.8,-2.9) {\bf b};
 \end{tikzpicture}
 \caption{({\bf a}) Equilibrium magnetization induced by the coupling between surface plasmon polariton fluctuations and the phonon mode $Q$, as a function of temperature $T$ and substrate-material distance $d$. The phonon energy is $\Omega = 13.5$ meV ($3.27$ THz), the spin-phonon coupling is $g = 0.5$ meV. The results were obtained from classical Monte Carlo simulations of the spin-phonon Hamiltonian in Eq.~\ref{eq:microscopic_ham} on a spin lattice with $20 \times 20$ unit cells (800 spins), at fixed coupling $\lambda$. The value for $\lambda$ was then translated into a distance $d$ as discussed in the text. Inset: The Raman phonon increases and decreases magnetic interactions within every other zig-zag chain, indicated by thick red lines and thin blue lines. This leads to a suppression of the fluctuations $\Delta S$ in the red chains, and an increase of $\Delta S$ in the blue chains, and thereby to a net magnetization. ({\bf b}) Line cuts of the data in {\bf a} for selected values of the substrate-material distance.}
 \label{fig:magnetization}
\end{figure}

To obtain the induced magnetization $\Delta M$, we can write $\Delta M = M_0 P_0$, where $P_0$ is the probability distribution for $M$ evaluated at $M_0$. Assuming the probability distribution is localized at the minimum, we can take $P_0 = 1$ to obtain $\Delta M = M_0 = c_M/\bar{a}_M$. Close to the critical temperature there is a finite magnetization
\begin{align}
 \Delta M = \Big( \frac{g\lambda}{\Omega L_s} \Big) \frac{\chi_L}{Va_M - (g^2/\Omega) \chi_L} \approx \Big( \frac{g\lambda}{\Omega a_M} \Big) \frac{\chi_L}{V L_s},
\end{align}
where we have assumed $Va_M \gg \chi_L$ in the macroscopic limit. This result is in good agreement with spin Monte Carlo simulations (see Fig.~\ref{fig:magnetization}), showing a finite magnetization developing for $T \approx T_{\rm N}$. A similar magnetization was recently observed when driving the system with intense THz radiation~\cite{Ilyas2023}.

From a finite size scaling analysis~\cite{Supplemental}, it is expected that $\chi_L \sim L^{\gamma/\nu} = L^{2-\eta}$ for temperatures close to $T_{\rm N}$, where $\eta$ is the anomalous dimension. Therefore,  the magnetization is expected to scale like $\chi_L/L^d \sim L^{-1/4}$ in two dimensions and $\chi_L/L^d \sim L^{-1.036}$ in three dimensions. However, these scaling laws assume a cubic system with volume $V = L^d$. In the present case, the system is highly asymmetric, with the vertical extent $L_z \ll L$. For such a system, the critical exponents are expected to follow the 3-D universality class, while the scaling of the susceptibility gives $\chi_L/V = (\chi_L/L^2) L_z^{-1} \sim L^{-0.036}$ in the macroscopic limit.

These scaling arguments combine a mean-field analysis of $F$ with renormalization group exponents for $\chi_L$ and $\xi$, and predict a magnetization weakly decaying with system size. This should be compared with our Monte Carlo simulations, which predict a size-independent magnetization density~\cite{Supplemental}. Since the Monte Carlo simulations include higher order terms in the Ginzburg-Landau theory, the discrepancy is likely due to higher order effects neglected in the analytical analysis. In particular, we note that using the mean-field exponents for $\chi_L$ gives an anomalous dimension $\eta = 0$, such that the magnetization becomes independent of volume close to $T_{\rm N}$, in line with the Monte Carlo results.


The results in Fig.~\ref{fig:magnetization} show that close to the N\'eel temperature, electric field fluctuations of the SPP modes induce a finite magnetization in FePS$_3$. In contrast to standard non-linear phononics, this magnetization appears in thermal equilibrium and thereby amounts to a true reconfiguration of the equilibrium magnetic state. Due to the strong links between non-linear phononics experiments and the present protocol, we expect that quite generally, non-equilibrium states of 2-D materials induced by resonant non-linear driving have an equilibrium analog realized through the coupling to SPPs. In particular, we note that the Ginzburg-Landau free energy discussed above is rather general, and appears across a wide range of systems with intertwined dominant and sub-dominant orders~\cite{Fradkin2015}. Provided that a coupling of the form $g LMQ$ is realized for some phonon mode, we therefore expect a similar phenomenology for the dominant and sub-dominant orders $L$ and $M$, irrespective of their physical nature. This observation opens the door to stabilizing exotic electronic phases in quantum materials in thermal equilibrium.

More broadly, we emphasize that while the particular effect of a phonon-induced magnetization is relatively moderate, the electric field fluctuations generated by the $\rm SrTiO_3$ substrate are very large in amplitude and rival the amplitude of state-of-the art THz lasers. This implies that equilibrium non-linear phononics could rival experiments performed by coherent lasers in a new noise driven context.


\begin{acknowledgments}
{\it Acknowledgments.--} We thank Daniel A. Bustamante Lopez and Dominik M. Juraschek for insightful discussions. EVB acknowledges funding from the European Union's Horizon Europe research and innovation programme under the Marie Sk{\l}odowska-Curie grant agreement No. 101106809 (CavityMag), and MHM acknowledges support from the Alexander von Humboldt Foundation. We further acknowledge support from the Cluster of Excellence ``CUI: Advanced Imaging of Matter''- EXC 2056 - project ID 390715994 and SFB-925 ``Light induced dynamics and control of correlated quantum systems'' – project 170620586 of the Deutsche Forschungsgemeinschaft (DFG), and Grupos Consolidados (IT1453-22), and from the Max Planck-New York City Center for Non-Equilibrium Quantum Phenomena at the the Flatiron Institute. The Flatiron Institute is a division of the Simons Foundation. 
\end{acknowledgments}


\bibliography{references}


\clearpage

\title{Supplemental material for "Equilibrium non-linear phononics by electric field fluctuations of terahertz cavities"}
\author{Emil \surname{Vi\~nas Bostr\"om}}

\thanks{These authors contributed equally to this work \\ \href{mailto:emil.bostrom@mpsd.mpg.de}{emil.bostrom@mpsd.mpg.de}}
\affiliation{Max Planck Institute for the Structure and Dynamics of Matter, Luruper Chaussee 149, 22761 Hamburg, Germany}
\affiliation{Nano-Bio Spectroscopy Group, Departamento de F\'isica de Materiales, Universidad del Pa\'is Vasco, 20018 San Sebastian, Spain}
\author{Marios H. Michael}
\thanks{These authors contributed equally to this work\\ \href{mailto:emil.bostrom@mpsd.mpg.de}{marios.michael@mpsd.mpg.de}}
\affiliation{Max Planck Institute for the Structure and Dynamics of Matter, Luruper Chaussee 149, 22761 Hamburg, Germany}
\author{Christian Eckhardt}
\affiliation{Max Planck Institute for the Structure and Dynamics of Matter, Luruper Chaussee 149, 22761 Hamburg, Germany}
\author{Angel Rubio}
\email{angel.rubio@mpsd.mpg.de}
\affiliation{Max Planck Institute for the Structure and Dynamics of Matter, Luruper Chaussee 149, 22761 Hamburg, Germany}
\affiliation{Center for Computational Quantum Physics, Flatiron Institute, Simons Foundation, New York City, NY 10010, USA}
\affiliation{Nano-Bio Spectroscopy Group, Departamento de F\'isica de Materiales, Universidad del Pa\'is Vasco, 20018 San Sebastian, Spain}

\date{\today}

\maketitle


\title{Supplemental material for "Equilibrium non-linear phononics by electric field fluctuations of terahertz cavities"}

\appendix
\renewcommand{\appendixname}{Supplementary Note}
\renewcommand{\figurename}{Supplementary Figure}
\setcounter{figure}{0}
\setcounter{equation}{0}

\section{Surface phonon polaritons}\label{app:spp}
The Hamiltonian of the surface phonon polariton (SPP) system is given by~\cite{Gubbin16}
\begin{align}
 H &= \frac{1}{2} \int_0^L dz\int d^2 x \, \bigg( \frac{{\bf D}^2}{\epsilon_0 \epsilon_{\infty} } + \mu_0 \left(\nabla \times {\bf A} \right)^2 - \frac{2\kappa {\bf X} \cdot {\bf D}}{\epsilon_0 \epsilon_{\infty}} \nonumber \\
 &\hspace*{2.8cm}+ \frac{{\bf P}^2}{\rho} + \rho \omega_{\rm LO}^2 {\bf X}^2 \bigg) \nonumber \\
 &+ \frac{1}{2} \int_{-L}^{0} dz\int d^2 x \, \bigg( \frac{{\bf D}^2}{ \epsilon_0} + \mu_0 \left(\nabla \times {\bf A} \right)^2  \bigg).
\end{align}
Here ${\bf D}$ is the displacement field, and is given through the relationship ${\bf D}({\bf r}) =  \epsilon_0 \epsilon_{\infty} {\bf E}({\bf r}) + \kappa {\bf X}$, with $\epsilon_0$ the vacuum permittivity and $\epsilon_\infty$ the dielectic constant of the substrate hosting the phonon polariton. Further, ${\bf A}$ is the electromagnetic vector potential, ${\bf X}$ is the shift of the ion coordinate, ${\bf P}$ is the corresponding ion momentum, and $\rho$ is the mass density of ions. The constant $\kappa^2 = \epsilon_0 \epsilon_{\infty} \rho \omega_{p}^2 $ is determined by the ion charge, through the phonon plasma frequency, $\omega_p$, and $\omega_{\rm TO}$ and $\omega_{\rm LO}^2 = \omega_{\rm TO}^2 + \omega_{p}^2$ are the  transverse and longitudinal optical phonon frequencies.
The Hamiltonian is written in the dipole gauge of the light-phonon system, but we note that we can use the dipole gauge for phonons and the Coulomb gauge for electrons without contradictions. Different gauges amount to unitary transformations that redefine which variable pairs act as generalized coordinates and momenta. 

In the dipole gauge the canonical commutation relations are given by
\begin{align}
    \left[ X_i({\bf r}), P_j({\bf r}')\right] =& i \delta_{ij} \delta({\bf r} - {\bf r}') \\
    \left[ D_i({\bf r}), A_j({\bf r}')\right] =& i \delta_{ij} \delta({\bf r} - {\bf r}'). \nonumber
\end{align}
As a function of position and frequency, Maxwell's equations lead to 
\begin{align}\label{eq:maxwell}
 -\frac{\omega^2\epsilon(\omega,z)}{c^2} {\bf A}({\bf r}) + \nabla \times [\nabla \times {\bf A}({\bf r})] = 0. 
\end{align}
Here the dielectric function is assumed to depend only on the vertical direction, with a dependence given by $\epsilon(\omega, z > 0) = \epsilon_{\infty} (\omega^2 - \omega_{\rm LO}^2)/(\omega^2 - \omega_{\rm TO}^2)$ and $\epsilon(\omega,z < 0 ) = 1$. Solutions of this equation lead to bulk and surface photon modes, which we label by ${\bf f}_n$ for the different mode functions, and by $\omega_n$ for the corresponding frequencies. The relations between ${\bf A}({\bf r},t)$ and the other coordinates are given by
\begin{align}
    {\bf E}({\bf r},\omega) &= i \omega {\bf A}({\bf r},\omega) \\
    {\bf X}({\bf r},\omega) &= - \frac{\kappa}{\omega^2 - \omega_{\rm TO}^2} {\bf E}({\bf r},\omega) = \sqrt{\frac{\epsilon_0}{\epsilon_{\infty} \rho}} \frac{\chi(\omega)}{\omega_{p}} {\bf E}({\bf r},\omega) \nonumber \\
    {\bf P}({\bf r},\omega) &= \rho(-i\omega) {\bf X}({\bf r},\omega), \nonumber
\end{align}
where the phonon susceptibility is defined as $\chi(\omega) = -(\epsilon_{\infty} \omega_{p}^2)/(\omega^2 - \omega_{\rm TO}^2)$. The mode functions satisfy the normalization conditions discussed in App.~\ref{app:normalization}.


\section{Mode function normalization}\label{app:normalization}
A straightforward way to compute the mode function normalization involves substituting the solution ${\bf f}_n(r)$ of Maxwell's equations, which gives the quantized vector potential $\mathbf{A}_n(r,\omega_m) = \mathbf{f}_n(r) e^{-i \omega_n t} \hat{a}_{n} + \mathbf{f}^*_{n}(r)e^{i \omega_n t}\hat{a}^\dag_{n}$, into the Hamiltonian, and requiring that it evaluates to the corresponding frequency. This gives
\begin{align}
    \omega_n ( \hat{a}_n \hat{a}^\dag_n + \hat{a}^\dag_n \hat{a}_n ) &= \bigg(\int_0^L dz\int d^2 x \, \epsilon_0 \omega_n^2 |{\bf f}_n(x,\omega_n)|^2 \nonumber \\
    &\times \bigg(\epsilon_{\infty} + \epsilon(\omega_n) + \frac{\omega_n^2 + \omega_{\rm TO}^2}{\epsilon_{\infty} \omega_{p}^2} \chi^2(\omega_n) \bigg)  \nonumber \\
    &+ \int_{-L}^{0} dz\int d^2 x \, 2\epsilon_0 \omega_n^2 |{\bf f}_n(x,\omega_n)|^2 \bigg) \nonumber \\
    &\times( \hat{a}_n \hat{a}^\dag_n + \hat{a}^\dag_n \hat{a}_n ),
\end{align}
implying the normalization condition
\begin{equation}
   \epsilon_0 \omega_n \int d^3 r \, v(\omega_n,{\bf r}) \epsilon(\omega_n,{\bf r}) |{\bf f}_n({\bf r},\omega_n)|^2 = 1.
\end{equation}
Here the weight function $v(\omega,{\bf r})$ is found in Ref.~\cite{Gubbin16} to be
\begin{equation}
    v(\omega) = \left( 2  + \frac{\omega}{\epsilon(\omega)} \frac{\partial[ \epsilon(\omega)]}{\partial \omega} \right),
\end{equation}
with $\epsilon(\omega,{\bf r})$ defined in the previous section. It can be verified through direct substitution that
\begin{align}
    &\int d^3 r \, v(\omega_n,{\bf r}) \epsilon(\omega_n,r) |{\bf f}_n({\bf r},\omega_n)|^2 \\
    &= \int d^3 r \left( \epsilon_{\infty} + \epsilon(\omega_n) + \frac{ \omega_n^2 + \omega_{\rm TO}^2}{\epsilon_{\infty} \omega_{p}^2} \chi^2(\omega_n) \right) |{\bf f}_n({\bf r},\omega_n)|^2. \nonumber
\end{align}


\section{Orthogonality}\label{app:orthogonality}
The Hamiltonian system is orthogonal with respect to the Poisson brackets, which have the schematic form 
\begin{align}
 \left\{ f_n , f^*_m \right\} = (-i) \int d^3 x [f_{nX}(x) f_{mP}^{*}(x) - f_{nP}(x) f_{mX}^{*}(x)].
\end{align}
Here $X$ and $P$ are different conjugate coordinate and momentum pairs. The resulting orthogonality relation is found to be
\begin{align}
 \delta_{nm} &= \epsilon_0 \int d^3 r \, f_n(x) f_m^*(x) \bigg( \omega_n \epsilon (\omega_n) + \omega_m \epsilon(\omega_m) \\
 &+ \frac{(\omega_m + \omega_n) \omega_n \omega_m\chi(\omega_m)  \chi(\omega_n)}{\epsilon_{\infty} \omega_{p}^2} \bigg) \nonumber
\end{align}
In fact, we can confirm that the Poisson bracket gives rise to the same normalization as the Hamiltonian approach when $\omega_n = \omega_m$, such that we have 
\begin{align}
   v(\omega_n) \epsilon(\omega_n) &= \epsilon_{\infty} + \epsilon(\omega_n) + \frac{ \omega_n^2 + \omega_{\rm TO}^2}{\epsilon_{\infty} \omega_{p}^2} \chi^2(\omega_n) \\
   &= 2 \epsilon(\omega_n) + 2 \frac{\omega_n^2}{\epsilon_{\infty} \omega_{p}^2} \chi^2(\omega_n). \nonumber
\end{align}


\section{Raman coupling and surface phonon polariton fluctuations}\label{app:raman}
The phonon mode at $3.27$ THz is Raman active and folded from the $M$-point of the paramagnetic unit cell. It therefore couples to the fluctuations of the SPP modes via a Raman-like term of the form
\begin{align}
 H_R &= \int d^3 x \frac{\epsilon_0 \hat{E}_l(x) \hat{E}_m(x)}{2} \frac{\partial \epsilon_{lm}({\bf Q}_i)}{\partial Q_i} u_i(x) \hat{Q}_i \\
 &\approx V_{\rm cell} \frac{\partial \epsilon_{lm}}{\partial Q_i} \frac{\hat{E}_l \hat{E}_m}{2} \hat{Q}_i, \nonumber
\end{align}
where in the last line all operators are evaluated in unit cell $i$. Here $u_i(x)$ corresponds to the wavefunction of the Raman mode at position $i$ which we take to be localised on a single unit-cell so that $\int d^3 x u_i(x) = V_{\rm cell}$.  In terms of phonon operators this coupling can be written as $H_R = \lambda (L/L_s)(b^\dagger + b)$, with the effective Raman coupling $\lambda = (\epsilon_0 V_{\rm c}l_0)/(2\sqrt{2}) \langle \hat{\bf E} \partial_Q\boldsymbol\epsilon \hat{\bf E} \rangle$. Here the factor $L/L_s$, with $L_{\rm s}$ the saturation magnetization, is included to account for the fact that the sign of $Q$ is defined relative to the sign of $L$~\cite{Ilyas2023}.

To estimate the value of $g$, we use SPPs to compute the local fluctuations $\langle E^2\rangle$. Expanding the modes as
\begin{align}
 E_{\parallel}({\bf x},z,t) &= \sum_q \frac{{\bf q} (- i \omega_s) f_q}{|q|} e^{- q z}e^{i {\bf q} \cdot {\bf x} - i \omega_s t}  \left( a^\dag_q - a_{-q} \right), \nonumber \\
 E_{\perp}({\bf x},z,t) &= \sum_q \omega_s f_q e^{- q z}e^{i {\bf q} \cdot {\bf x} - i \omega_s t}  \left( a^\dag_q + a_{-q} \right) \\
 f_q &= \sqrt{\frac{q (\omega_s^2 - \omega_{\rm TO}^2)}{4 A \epsilon_0 (\epsilon_{\rm sub} + \epsilon_{\rm mat}) \omega_s^3}}, \nonumber
\end{align}
where $\omega_s$ is the SPP frequency, $A$ the surface area, $\omega_{\rm TO}$ the frequency of the transverse optical (TO) phonon, $\epsilon_{\rm sub}$ and $\epsilon_{\rm mat}$ the relative dielectric constants of the phonon-polariton substrate and the material, respectively, and $d$ the substrate-material distance. The electric field fluctuations can be evaluated like
\begin{align}
 \epsilon_0 \sum_{lm} \langle E_l E_m \rangle &= \sum_{lm} \delta_{lm} \left( \frac{\delta_{lx}}{2} + \frac{\delta_{ly}}{2} + \delta_{lz} \right)  \\
 &\times \int \frac{d^2q}{(2\pi)^2} \frac{(\omega_s^2 - \omega_{\rm TO}^2)}{4 (\epsilon_{\rm sub} + \epsilon_{\rm mat}) \omega_s} q e^{-2 q d} \nonumber \\
 &= 2 \times \frac{\omega_s^2 - \omega_{\rm TO}^2}{8\pi \omega_{s}(\epsilon_{\rm sub} + \epsilon_{\rm mat})} \int_{0}^{\infty} q^2 e^{-2 qd} dq \nonumber \\
 &= \frac{\omega_s^2 - \omega_{\rm TO}^2}{16 \pi \omega_{s}(\epsilon_{\rm sub} + \epsilon_{\rm mat})d^3} \nonumber
\end{align}
For a SrTiO$_3$ substrate a distance $5$ nm away from the magnetic material, this gives electric field fluctuations of the order $\langle E^2\rangle \sim 1$ MV m$^{-1}$. We note that this has the general form of a characteristic frequency ($\omega_s$) over a characteristic volume ($d^3$).


\section{Fluctuations of a Fabry-Perot cavity}\label{app:fp_cavity}
For comparison, we perform the analogous calculation for a Fabry-Perot cavity. As an example, the electric field fluctuations parallel to the two mirrors in the center of the cavity can be obtained from
\begin{align}
 \epsilon_0 \langle E_{\parallel}^2(z_{c}) \rangle &- \epsilon_0 \langle E_{\parallel, \rm free}^2 \rangle = \int d\omega \frac{\omega}{2} [\rho(\omega,z_{c}) - \rho_0(\omega)],
\end{align}
where $\rho(\omega,z_{c})$ is the Fabry-Perot photonic density of states (DOS), $z_c$ is the center of the cavity in the vertical direction (half-way between the two mirrors), and $\rho_0 = \omega^2/(3\pi c^3)$ is the DOS of free space. To a good approximation, the ratio $\rho/\rho_0$ is given by the expression~\cite{Eckhardt2024, Chris_arXiv}
\begin{align}
 \frac{\rho}{\rho_0} = \frac{2\omega_c}{\omega} \sum_{n=0}^\infty \theta[\omega - (2n + 1)\omega_c],
\end{align}
where $\theta(\omega)$ is the Heaviside distribution. Using this approximation, we can write the electric field fluctuations as
\begin{align}
\epsilon_0 \langle E^2 \rangle &- \epsilon_0 \langle E_{\rm free}^2 \rangle = \int d\omega \frac{\omega^3}{6\pi c^3} \bigg[\frac{\rho(\omega)}{\rho_0(\omega)} - 1 \bigg] \\
 &= \frac{\omega_c}{6\pi c^3} \int d\omega \omega^2 \bigg[ 2\sum_{n=0}^\infty \theta[\omega - (2n + 1)\omega_c] - \frac{\omega}{\omega_c} \bigg]. \nonumber 
\end{align}
We now use the fact that the function inside the square brackets is a saw-tooth function of period $2\omega_c$, and write the fluctuations as
\begin{align}
 \epsilon_0 \langle E^2 \rangle &- \epsilon_0 \langle E_{\rm free}^2 \rangle = \frac{\omega_c}{6\pi c^3} \int d\omega \omega^2 e^{-\eta\omega} \sum_{n\neq 0} ic_n e^{i\pi n\omega/\omega_c} \nonumber \\
 &= \frac{\omega_c}{6\pi c^3} \int d\omega \frac{\partial^2}{\partial\eta^2} \sum_{n\neq 0} ic_n e^{i\pi n\omega/\omega_c} e^{-\eta\omega} \\
 &= \frac{\omega_c}{3\pi c^3} \sum_{n\neq 0} \frac{-ic_n}{(i\pi n/\omega_c - \eta)^3} \nonumber \\
 &= \frac{\omega_c}{3\pi d^3} \sum_{n\neq 0} \frac{c_n}{n^3}. \nonumber
\end{align}
Here we have introduced the factor $e^{-\eta\omega}$ to regularize the integral as $\omega \to \infty$, where $\eta$ is a positive infinitesimal, and defined $d = c\pi/\omega_c$ as the height of the cavity. 

It now remains to evaluate the sum. For a saw-tooth function we have $c_n = (-1)^{n-1}/(\pi n)$ and $c_{-n} = c_n$, such that the expression above becomes
\begin{align}
 \epsilon_0 \langle E^2 \rangle &- \epsilon_0 \langle E_{\rm free}^2 \rangle = \frac{2\omega_c}{3\pi^2 d^3} \sum_{n>0} \frac{(-1)^{n-1}}{n^4} \\
 &= \frac{2\omega_c}{3\pi^2 d^3} \eta(4). \nonumber
\end{align}
Here $\eta(n)$ is the Dirichlet $\eta$-function, which for $n = 4$ is $\eta(4) = 7\pi^4/720 \approx 0.95$. Using this result, we find
\begin{align}
 \epsilon_0 \langle E^2 \rangle &- \epsilon_0 \langle E_{\rm free}^2 \rangle = \frac{7\pi^2\omega_c}{1080d^3}.
\end{align}

We note that again this expression has the general form of a characteristic frequency ($\omega_c$) over a characteristic volume ($d^3$). The difference compared to the SPPs is that for the Fabry-Perot cavity the characteristic length scale $d$ and the frequency $\omega_c$ are proportional, such that a small mode volume requires a large frequency. In contrast, for the SPPs, $d$ and $\omega_s$ are independent quantities.



\section{Phonon-polariton density of states}
We now calculate the frequency resolved density of states of the SPPs, defined as
\begin{align}
 \nu(\omega) = \sum_{\bf q} \delta(\omega - \omega_{\bf q}) = \frac{1}{4\pi^2} \int d\phi \int dq q\, \delta(\omega - \omega_{\bf q}).
\end{align}
Here, the second equality holds in the macroscopic limit $L \to \infty$. We note that the DOS, $\nu(\omega)$ merely discusses properties of the dispersion of the SPP modes, not to be confused with the photonic local density of states, $\rho(\omega,\vec{x})$, discussed in the previous section which also contains information about the overlap of each mode with light.

The phonon-polariton dispersion is
\begin{align}
 2\omega_{\bf q}^2 = 2 c^2 q^2 + \omega_{\rm LO}^2 - \sqrt{\omega_{\rm LO}^4 - 4 c^2 q^2 \omega_{\rm TO}^2 + 4 c^4 q^4},
\end{align}
which in the large $q$ limit becomes $\omega_s^2 = (\omega_{\rm LO}^2 + \omega_{\rm TO}^2)/2$. If we define the variable $\bar{\omega} = cq$, the integral becomes
\begin{align}
 \nu(\omega) = \frac{1}{2\pi c^2} \int d\bar{\omega} \bar{\omega}\, \delta[\omega - \omega_{\bf q}(\bar{\omega})].
\end{align}
To evaluate the integral we now make the substitution $\omega'(\bar{\omega}) = \omega_{\bf q}(\bar{\omega})$, from which we can obtain the density of states as
\begin{align}
 \nu(\omega) &= \frac{1}{2\pi c^2} \int d\bar{\omega} \bar{\omega}\, \delta[\omega - \omega'(\bar{\omega})] \\
 &= \frac{1}{2\pi c^2} \int d\omega' \bigg[ \bar{\omega} \frac{\partial\bar{\omega}}{\partial\omega'} \bigg](\omega') \delta(\omega - \omega') \nonumber \\
 &= \frac{1}{2\pi c^2} \bigg( \bar{\omega} \frac{\partial\bar{\omega}}{\partial\omega'} \bigg) \bigg|_{\omega'=\omega}. \nonumber
\end{align}
Here, in the second line, the expression in brackets is to be considered as a function of $\omega'$.

To evaluate this expression, we first find the inverse relation $\bar{\omega}(\omega')$, that can be shown to be
\begin{align}
 \bar{\omega} = \omega' \frac{\sqrt{\omega_{\rm LO}^2 - \omega'^2}}{\sqrt{\omega_{\rm LO}^2 + \omega_{\rm TO}^2 - 2\omega'^2}}.
\end{align}
By straightforward differentiation we then find
\begin{align}
 \nu(\omega) &= \frac{1}{2\pi c^2} \bar{\omega} \frac{\partial\bar{\omega}}{\partial\omega'} \bigg|_{\omega'=\omega} \\
 &= \frac{\omega}{2\pi c^2} \frac{\omega_{\rm LO}^2 (\omega_{\rm LO}^2 + \omega_{\rm TO}^2 - 2\omega^2) + 2\omega^2 (\omega^2 - \omega_{\rm TO}^2)}{(\omega_{\rm LO}^2 + \omega_{\rm TO}^2 - 2\omega^2)^2}. \nonumber
\end{align}


\section{Derivation of the local Ginzburg-Landau theory}\label{app:gl_derivation}
The low-energy magneto-elastic properties of FePS$_3$ can be described by a Hamiltonian of the form \cite{Cui2023,Liu2021}
\begin{align}\label{eq:ham}
 H &= \sum_{ij} \bS_i \cdot ({\bf J}_{ij} \bS_j) - \Delta \sum_i (S_i^z)^2 \\
 &+ \frac{1}{2} \sum_\alpha \bigg[ P_\alpha^2 + \Omega_\alpha^2 Q_\alpha^2 \bigg], \nonumber
\end{align}
where the matrix ${\bf J}_{ij}$ encodes the (possibly anisotropic) couplings of spins $\bS_i$ and $\bS_j$, and $\Delta$ is a single ion anisotropy. $P_\alpha$ and $Q_\alpha$ are the momentum and displacement operators of phonon mode $\alpha$, respectively, and $\Omega_\alpha$ is the corresponding phonon frequency. In general, the magnetic interactions depend on the atomic positions, giving rise to an interaction between the spins and the lattice displacements. For small displacements $Q_\alpha$ the exchange tensor can be expanded as
\begin{align}\label{eq:exchange} 
 {\bf J}_{ij}(Q) \approx {\bf J}_{ij} - \boldsymbol\alpha_{ij} Q,
\end{align}
where the tensor $\boldsymbol\alpha_{ij}$ quantifies the phonon modulation of the magnetic interactions.

The magnetic exchange interactions as well as the modulations $\boldsymbol\alpha_{ij}$ for the $3.27$ THz phonon mode were calculated from first principles within the frozen phonon approximation~\cite{Ilyas2023}. We find that $\boldsymbol\alpha_{ij}$ decays rapidly with inter-atomic distance, and is significant only for nearest neighbor exchange interactions. Due to the space group symmetries of the system, only six nearest neighbor bonds of the magnetic unit cell are independent, and are henceforth labeled $J_i$ in accordance with Fig.~\ref{fig:phonon_patterns}.

The spatial pattern in which the phonon modulates the magnetic interactions is shown in Fig.~\ref{fig:phonon_patterns}, and the strength of the couplings was found to be $|\alpha| = 17$ meV \AA$^{-1}$. As the diagonal components of $\alpha$ are much larger than the off-diagonal components, $\alpha$ can be treated as a scalar. 

\begin{figure}[b]
\centering
    \begin{tikzpicture}
    \newdimen\R
    \newdimen\Dx
    \newdimen\Dy
    \newdimen\Tx
    \newdimen\Ty
    \R=0.7cm
    \Tx=1.732050875688772\R                 
    \Ty=0.8660254037844386\R  
    \Dx=-0.5\Tx
    \Dy=1.732050875688772\Ty
            
    \begin{scope}[shift={(0.0, 0.0)},rotate=0,scale=1.4]
      \draw[line width=0.10mm] (-30:\R) \foreach \x in { 30} { -- (\x:\R)};
      \draw[line width=0.10mm] (150:\R) \foreach \x in {210} { -- (\x:\R)};
      \draw[line width=0.20mm, color=blue!60!white] ( 30:\R) \foreach \x in { 90,150} { -- (\x:\R)};
      \draw[line width=0.50mm, color=red!60!white ] (210:\R) \foreach \x in {270,330} { -- (\x:\R)};
      \foreach \x in {30,90,150}
      \node[inner sep=2.0pt, circle, draw, color=blue!80!white, fill=blue!80!white] at (\x:\R) {};
      \foreach \x in {210,270,330}
      \node[inner sep=2.0pt, circle, draw, color=red!80!white , fill=red!80!white ] at (\x:\R) {};
    \end{scope}
    
    \begin{scope}[shift={(1.4*\Tx, 0.0)},rotate=0,scale=1.4]
      \draw[line width=0.10mm] (-30:\R) \foreach \x in { 30} { -- (\x:\R)};
      \draw[line width=0.10mm] (150:\R) \foreach \x in {210} { -- (\x:\R)};
      \draw[line width=0.20mm, color=blue!60!white] ( 30:\R) \foreach \x in { 90,150} { -- (\x:\R)};
      \draw[line width=0.50mm, color=red!60!white ] (210:\R) \foreach \x in {270,330} { -- (\x:\R)};
      \foreach \x in {30,90,150}
      \node[inner sep=2.0pt, circle, draw, color=blue!80!white, fill=blue!80!white] at (\x:\R) {};
      \foreach \x in {210,270,330}
      \node[inner sep=2.0pt, circle, draw, color=red!80!white , fill=red!80!white ] at (\x:\R) {};
    \end{scope}
     
    \begin{scope}[shift={(1.4*\Dx, 1.4*\Dy)},rotate=0,scale=1.4]
      \draw[line width=0.10mm] (-30:\R) \foreach \x in { 30} { -- (\x:\R)};
      \draw[line width=0.10mm] (150:\R) \foreach \x in {210} { -- (\x:\R)};
      \draw[line width=0.50mm, color=red!60!white] ( 30:\R) \foreach \x in { 90,150} { -- (\x:\R)};
      \draw[line width=0.20mm, color=blue!60!white ] (210:\R) \foreach \x in {270,330} { -- (\x:\R)};
      \foreach \x in {30,90,150}
      \node[inner sep=2.0pt, circle, draw, color=red!80!white, fill=red!80!white] at (\x:\R) {};
      \foreach \x in {210,270,330}
      \node[inner sep=2.0pt, circle, draw, color=blue!80!white , fill=blue!80!white ] at (\x:\R) {};
    \end{scope}

    \begin{scope}[shift={(1.4*\Dx+1.4*\Tx, 1.4*\Dy)},rotate=0,scale=1.4]
      \draw[line width=0.10mm] (-30:\R) \foreach \x in { 30} { -- (\x:\R)};
      \draw[line width=0.10mm] (150:\R) \foreach \x in {210} { -- (\x:\R)};
      \draw[line width=0.50mm, color=red!60!white] ( 30:\R) \foreach \x in { 90,150} { -- (\x:\R)};
      \draw[line width=0.20mm, color=blue!60!white ] (210:\R) \foreach \x in {270,330} { -- (\x:\R)};
      \foreach \x in {30,90,150}
      \node[inner sep=2.0pt, circle, draw, color=red!80!white, fill=red!80!white] at (\x:\R) {};
      \foreach \x in {210,270,330}
      \node[inner sep=2.0pt, circle, draw, color=blue!80!white , fill=blue!80!white ] at (\x:\R) {};
    \end{scope}
     
     \node at (1.40-2,-2.45+1.4) {$J_1 \uparrow$};
     \node at (1.40-2,-0.40+1.4) {$J_2 \downarrow$};
     \node at (0.85-2,-1.40+1.4) {$J_3$};
     \node at (2.30-2, 0.10+1.4) {$J_4$};
     \node at (2.60-2,-2.45+1.4) {$J_5 \uparrow$};
     \node at (2.60-2,-0.40+1.4) {$J_6 \downarrow$};

     \node at (1.15-0,-1.70+1.4) {$S_1$};
     \node at (1.15-0,-1.10+1.4) {$S_2$};
     \node at (1.70-0,-0.80+1.4) {$S_3$};
     \node at (1.70-0,-2.00+1.4) {$S_4$};
    
  \end{tikzpicture}
 \caption{Spatial modulation pattern of the magnetic interactions induced by the $3.27$ THz Raman phonon mode. An increase in the magnetic interaction on a given bond is illustrated by a thick line and an upward-pointing arrow $\uparrow$, while a decrease is illustrated by a thin line and a downward-pointing arrow $\downarrow$.}
 \label{fig:phonon_patterns}
\end{figure}

Defining the zigzag (ZZ) antiferromagnetic and ferromagnetic (FM) order parameters $L = S_1 - S_2 - S_3 + S_4$ and $M = S_1 + S_2 + S_3 + S_4$ (within a given magnetic unit cell), the microscopic modulations of the magnetic parameters can be mapped onto a theory of macroscopic variables. These macroscopic variables are expected to govern the low-energy and long wavelength behavior of the theory. More specifically, we can write the spin-phonon coupling as
\begin{align}
 H_{\rm int}^{(i)} &= \gamma Q_i (S_{2i} S_{3i} - S_{1i} S_{4i}) \nonumber \\
 &= -g Q_i L_i M_i
\end{align}
where $Q_i$ is the Raman phonon coordinate, $L_i$ the zigzag order and $M_i$ the magnetization of cell $i$, coupled through the constant $g$.


\section{Long wavelength Ginzburg-Landau theory}\label{app:gl_theory}
In the macroscopic limit, we can transform the local theory discussed above into a long wavelength theory in terms of reduced density variables. Defining the Fourier transforms
\begin{align}
 L_i &= \frac{1}{N} \sum_q e^{iqr_i} L_q \approx \frac{L_0}{N} = L\\
 M_i &= \frac{1}{N} \sum_q e^{iqr_i} M_q \approx \frac{M_0}{N} = M\\
 Q_i &= \frac{1}{N} \sum_q e^{iqr_i} Q_q \approx \frac{Q_0}{N} = Q,
\end{align}
we will use $L$, $M$ and $Q$ to denote the density of the zero momentum component of each variable. 

Starting from a local Ginzburg-Landau theory of the form
\begin{align}\label{eq:ginzburg_landau}
 F &= \sum_i \Big[ \frac{a_L(T)}{2} L_i^2 + \frac{b_L}{4} L_i^4 + \frac{a_M(T)}{2} M_i^2 + \frac{\Omega}{2} Q_i^2 \nonumber \\
 &+ g (L_i -\langle L_i\rangle) M_i Q_i \Big],
\end{align}
it is then straightforward to map the local theory onto an effective theory formulated in terms of the variables $L$, $M$ and $Q$. In this effective theory, the system is described by the Ginzburg-Landau free energy
\begin{align}
 F &= \sum_i \Big[ \frac{a_L(T)}{2} L^2 + \frac{b_L}{4} L^4 + \frac{a_M(T)}{2} M^2 + \frac{\Omega}{2} Q^2 \nonumber \\
 &+ g (L -\langle L\rangle) M Q \Big]. \nonumber \\
 &= V\Big[ \frac{a_L(T)}{2} L^2 + \frac{b_L}{4} L^4 + \frac{a_M(T)}{2} M^2 + \frac{\Omega}{2} Q^2 \\
 &+ g (L -\langle L\rangle) M Q \Big]. \nonumber
\end{align}
Since all (reduced) variables are defined as densities, the overall prefactor $V$ ensures that the free energy is extensive. Measuring all parameters in units of the temperature we have the free energy 
\begin{align}
 F &= TV \Big[ \frac{a_L(T)}{2} L^2 + \frac{b_L}{4} L^4 + \frac{a_M(T)}{2} M^2 + \frac{\Omega}{2} Q^2 \\
 &+ g (L -\langle L\rangle) M Q \Big]. \nonumber
\end{align}


\section{Fitting the Ginzburg-Landau theory to Monte Carlo data}\label{app:mc_fit}
\begin{figure}
    \centering
    \begin{tikzpicture}
     \node at ( 0, 6.4) {\includegraphics[width=\columnwidth]{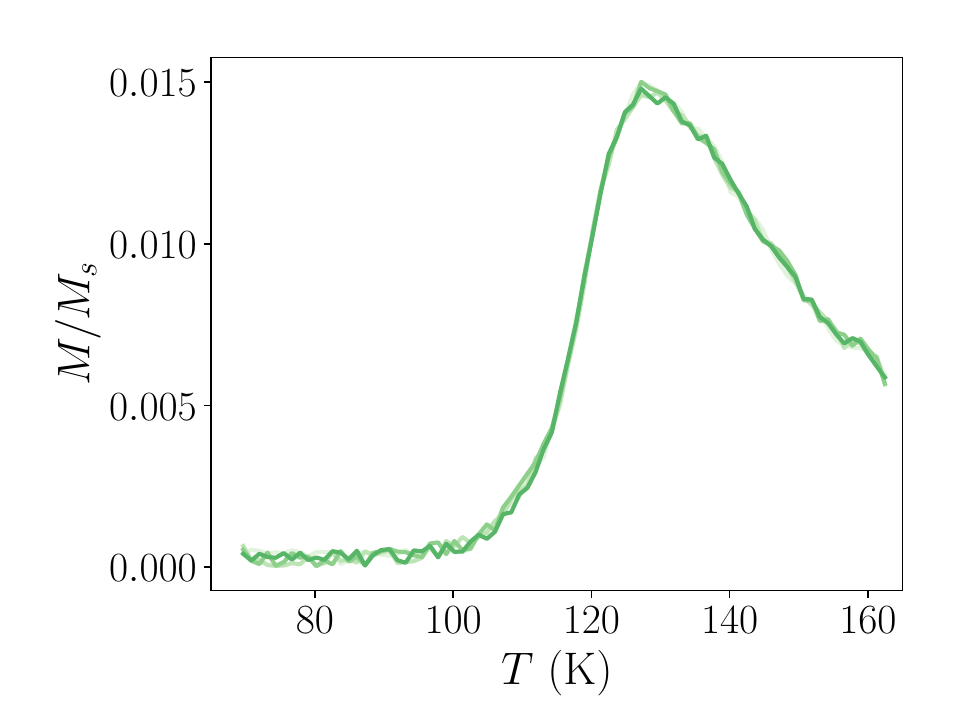}};
     \node at ( 0, 0.0) {\includegraphics[width=\columnwidth]{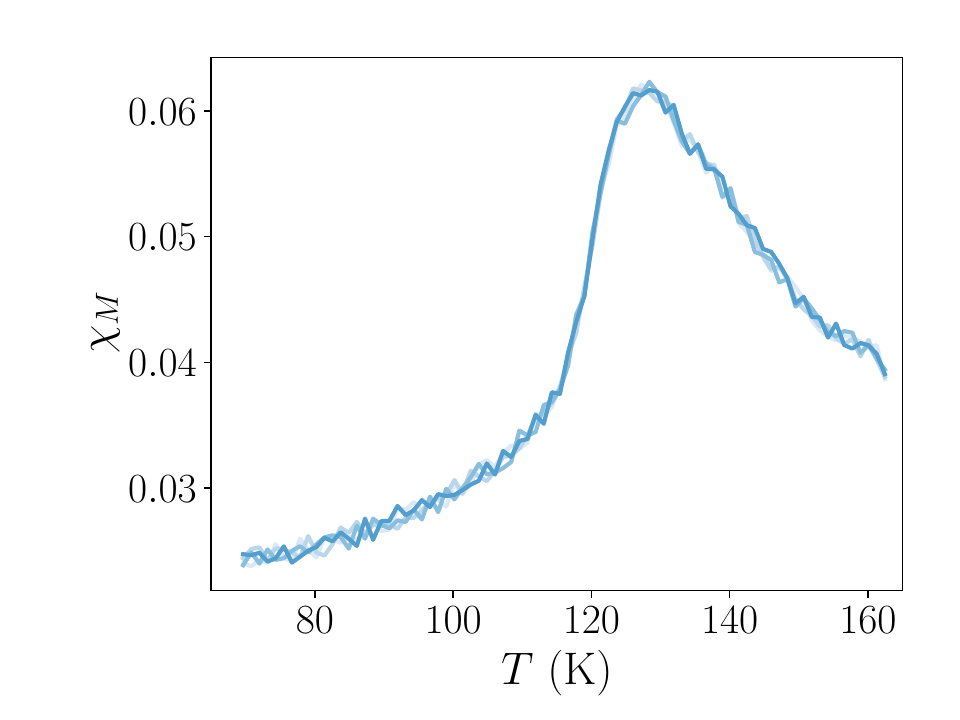}};
     \node at ( 0,-6.4) {\includegraphics[width=\columnwidth]{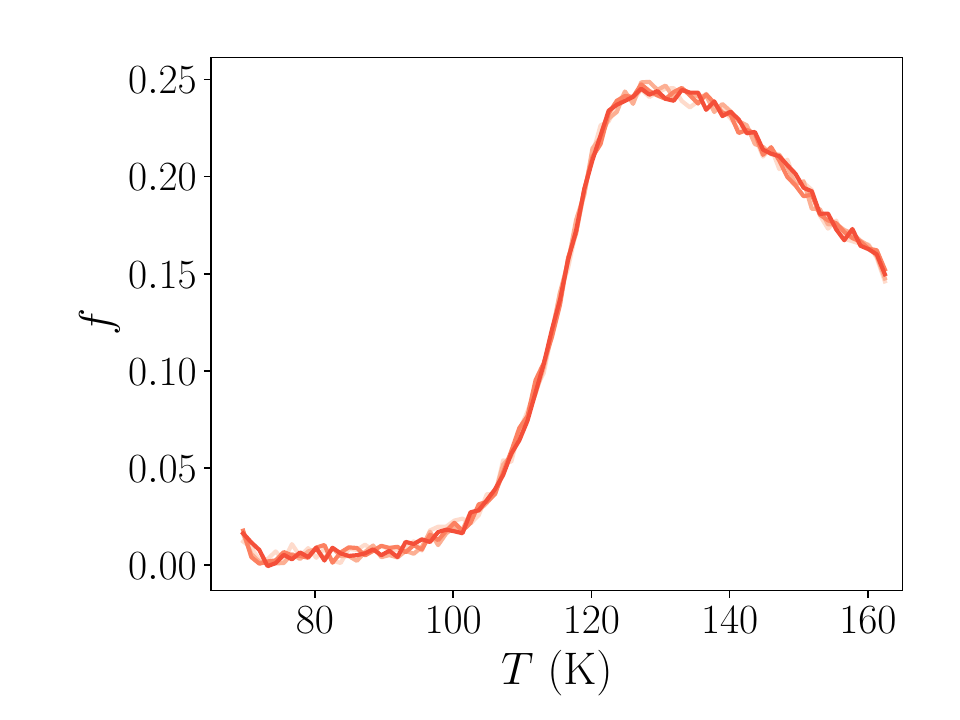}};
     \node at (-2, 8.4) {\Large\bf a};
     \node at (-2, 2.0) {\Large\bf b};
     \node at (-2,-4.4) {\Large\bf c};
    \end{tikzpicture}
    \caption{Ferromagnetic magnetization ({\bf a}) and susceptibility ({\bf b}) calculated with spin Monte Carlo for fixed phonon displacement $Q$ and different systems of linear size $L = 20$, $24$, $28$ and $32$ (colored from light to dark). The data of panels {\bf a} and {\bf b} is used to numerically estimate the coupling between $M$ and $Q$, using the relation $f = \langle M\rangle/\chi_M$ ({\bf c}).}
    \label{fig:mc_data}
\end{figure}

To obtain an effective theory for the magnetization, we assume a Ginzburg-Landau free energy of the form
\begin{align}
 F(M,T) &= \frac{a}{2} M^2 + f M.
\end{align}
Defining the partition function $Z = \int dM e^{-F/T}$, and the $n$th moment of the magnetization via $\langle M^n\rangle = Z^{-1} \int dM M^n e^{-F/T}$, we have
\begin{align}
 \langle M\rangle &= \frac{f}{a} \\
 \langle M^2\rangle &= \frac{f^2}{a^2} + \frac{T}{a}.
\end{align}
These equations have the solution
\begin{align}
 a &= \frac{T}{\langle M^2\rangle - \langle M\rangle^2} = \frac{TV}{\chi_M} \\
 f &= \frac{T\langle M\rangle}{\langle M^2\rangle - \langle M\rangle^2} = \frac{TV \langle M\rangle}{\chi_M}.
\end{align}
Since $\chi_M = V (\langle M^2\rangle - \langle M\rangle^2)$ is independent of volume, both these parameters will increase with system size as $V = L^2$. The magnetization $\langle M\rangle$ is however independent of system size, as also verified by Monte Carlo data. To keep the explicit temperature and system size dependence transparent, we write 
\begin{align}
 F(M,T) &= TV \big[ \frac{a}{2} M^2 + f M \big] \\
 a &= \frac{1}{\chi_M} \\
 f &= \frac{\langle M\rangle}{\chi_M}.
\end{align}
The dimensionless parameters $a$ and $f$ can easily be parameterized by calculating $\langle M\rangle$ and $\chi_M$ from spin Monte Carlo, as shown in Fig.~\ref{fig:mc_data}.

\section{Self-consistent spin-phonon Monte Carlo scheme}\label{app:sc_mc}
To obtain the temperature dependent magnetization resulting from a finite phonon-SPP and spin-phonon coupling, we utilize a self-consistent Monte Carlo scheme.

\begin{figure}
    \centering
    \begin{tikzpicture}
     \node at ( 0, 3.2) {\includegraphics[width=\columnwidth]{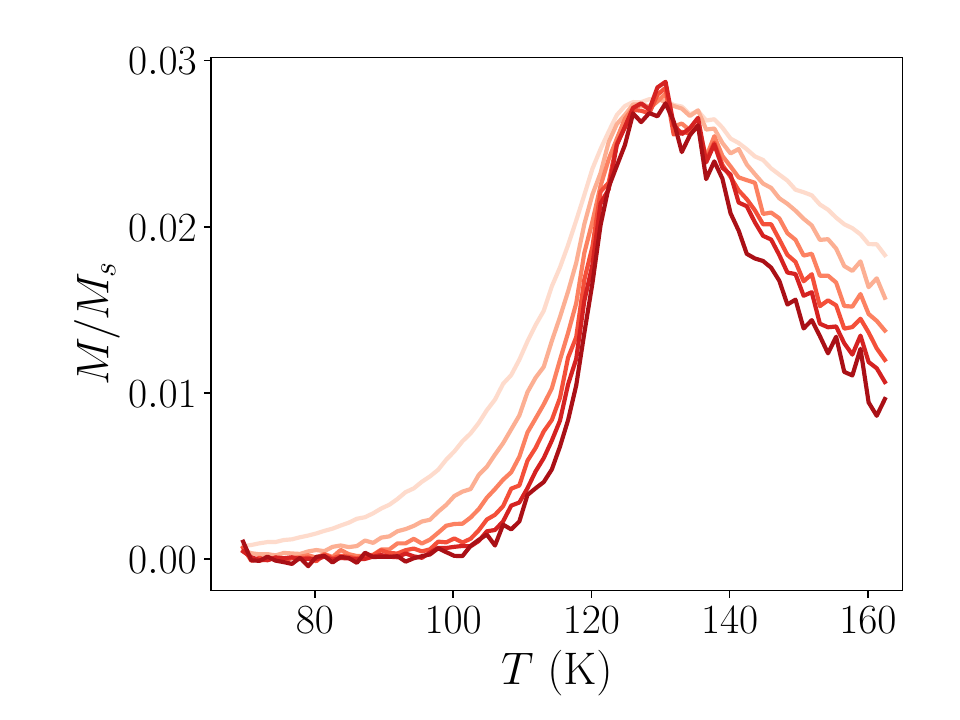}};
     \node at ( 0,-3.2) {\includegraphics[width=\columnwidth]{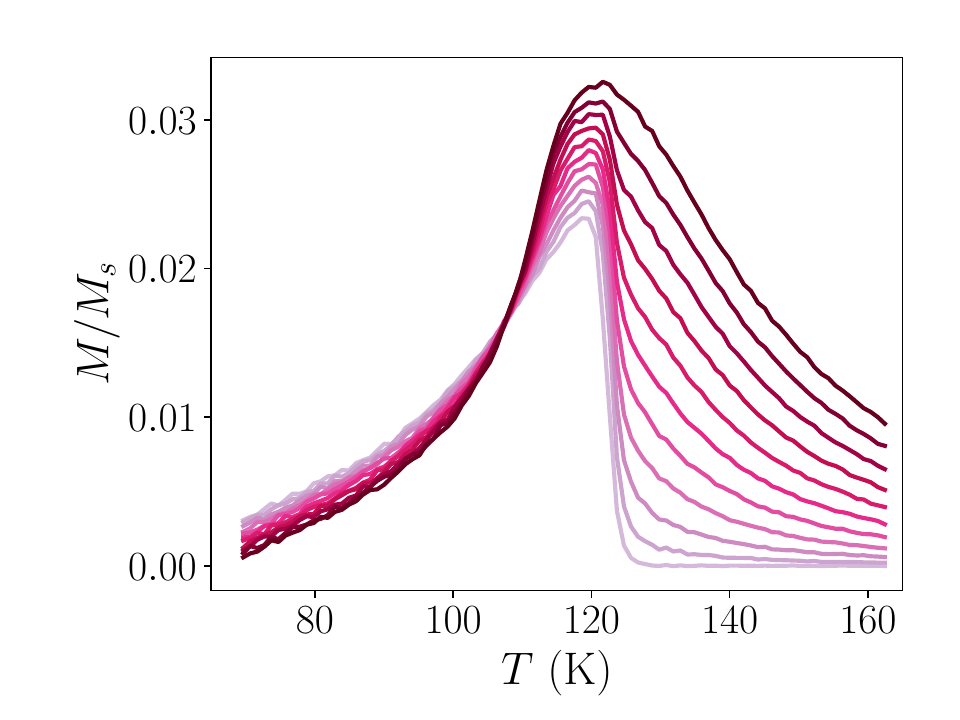}};
     \node at (-2, 5.2) {\Large\bf a};
     \node at (-2,-1.2) {\Large\bf b};
    \end{tikzpicture}
    \caption{({\bf a}) Temperature dependent magnetization obtained by the self-consistent Monte Carlo scheme discussed below Eq.~\ref{eq:sc_mc}. The different curves show the magnetization resulting from six iterations, starting with light ($n = 1$) and ending with dark ($n = 6$) lines. ({\bf b}) Temperature dependence of the SPP and phonon induced magnetization, obtained from Monte Carlo simulations with different range of the cut-off used to evaluate the local order parameter $L_i$. The dark red line corresponds to the region $R_i$ consisting of nearest neighbors while the light purple line corresponds to $R_i$ being the entire lattice. For the lines in between ($n = 2, 3, \ldots, 10$), $R_i$ is a circle with radius $R = na$ and $a$ is the nearest neighbor distance.}
    \label{fig:mc_scalings}
\end{figure}

The total spin-phonon Hamiltonian, containing the phonon-SPP coupling, is given by
\begin{align}
 H_{\rm s-ph} &= H_{\rm s} + H_{\rm ph} - g \sum_{ij} Q_{i} \alpha_{ij} {\bf S}_i \cdot {\bf S}_j \\
 &- \lambda \sum_{i\in R_i} \frac{L_i}{L_s} Q_i. \nonumber
\end{align}
Since there is no straightforward way to consider $Q_i$ as dynamical variables within the Monte Carlo simulation, we approximate the solution to the above Hamiltonian with an effective description in terms of the coupled equations
\begin{align}\label{eq:sc_mc}
 H_{\rm s, eff} &= H_{\rm s} - g \sum_{ij} Q_{i}(T) \alpha_{ij} {\bf S}_i \cdot {\bf S}_j \\
 Q_{i}(T) &= \frac{1}{\Omega} \Big( g \sum_j \alpha_{ij} {\bf S}_i \cdot {\bf S}_j + \lambda \sum_{i\in R_i} \frac{L_i}{L_s} \Big) \nonumber \\
 &= Q_{\rm s-ph}(T) + \frac{\lambda}{\Omega} \sum_{i\in R_i} \frac{L_i}{L_s}.
\end{align}
Here, the function $Q_{\rm s-ph}(T)$ is taken to be constant during a given Monte Carlo simulation, and is calculated for iteration $n$ from the spin correlations functions of iteration $n-1$. The magnetization obtained by self-consistent iteration of these equations is shown in Fig.~\ref{fig:mc_scalings}a. The last term, coming from the phonon-SPP coupling, gives a constant (but temperature dependent) force acting on the phonon coordinate.

A technical point related to the phonon-SPP coupling is the range $R_i$ over which to evaluate the local order parameter $L_i$. Depending on the choice of $R_i$, the high temperature tail of the magnetization goes all the way from a slow linear decay away from $T_{\rm N}$ (for a small, nearest neighbor, sampling region), to a sharp order parameter like onset at $T_{\rm N}$ (when $L_i$ is sampled over the entire lattice). This behavior can be understood by noting that local order starts to form already far above $T_{\rm N}$, while long range order only appears at $T_{\rm N}$. In Fig.~\ref{fig:mc_scalings}b we show the behavior of the magnetization for different values of $R_i$, ranging from the local (nearest neighbor) limit to the long range limit. 

In Fig.~\ref{fig:mc_scalings} the range $R_i$ is kept fixed during the simulations, while a more realistic scheme should consider a temperature dependent $R_i$. Since the average length over which the lattice is ordered is given by the correlation length $\xi$, which grows polynomially as $T_{\rm N}$ is approached from above, a possible functional form would be to take $R_i(T)$ as a circle centered at site $i$ with radius $\xi(T)$.

\section{Dependence of magnetization on cavity-phonon and spin-phonon coupling}\label{app:lambda}
\begin{figure}
    \centering
    \begin{tikzpicture}
     \node at ( 0, 3.2) {\includegraphics[width=\columnwidth]{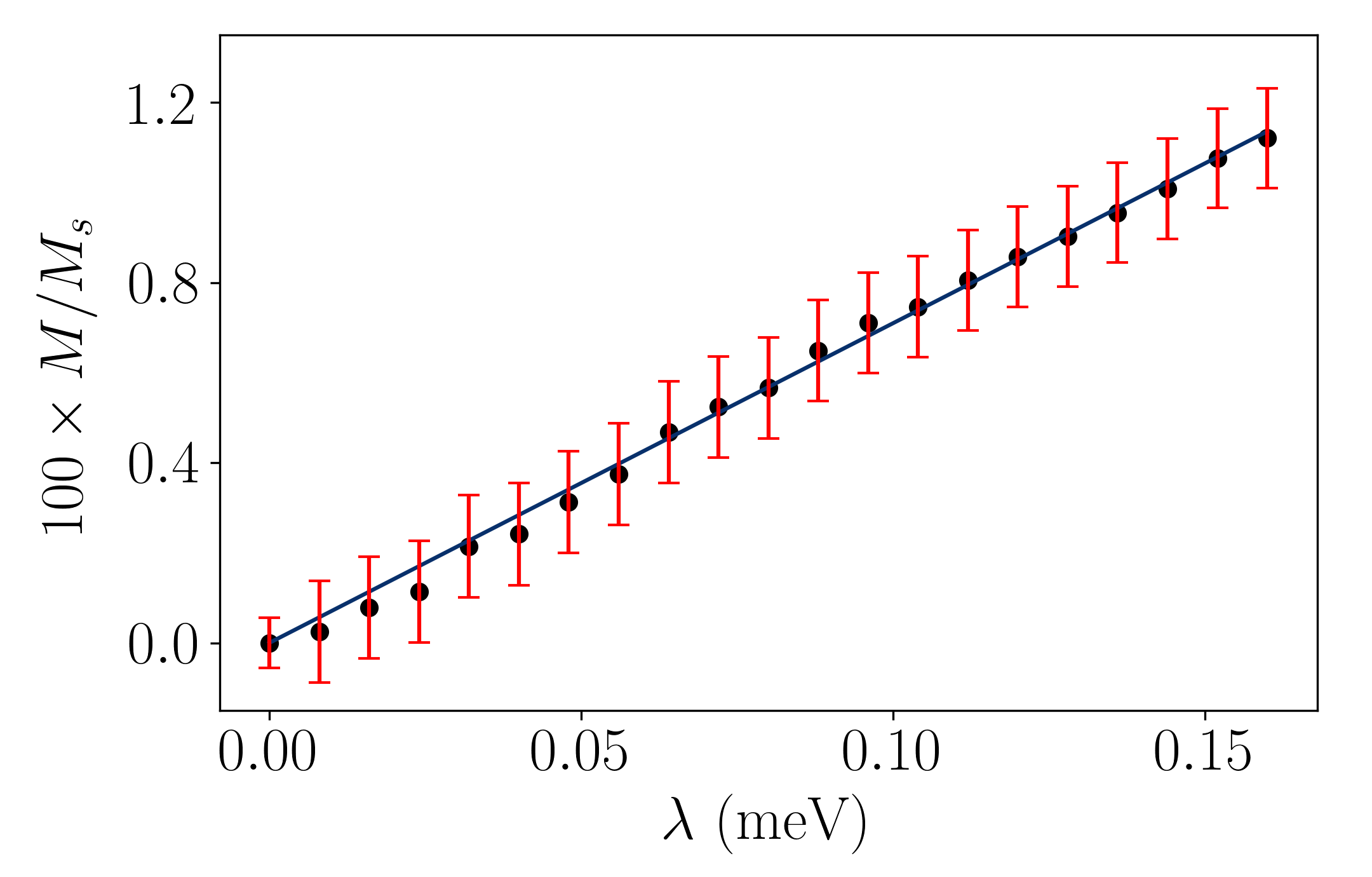}};
    \end{tikzpicture}
    \caption{Dependence of the induced magnetization on the Raman-cavity coupling $\lambda$, at the N\'eel temperature $T = T_{\rm N}$. The results were obtained via the self-consistent Monte Carlo scheme described in Supplementary Section~\ref{app:sc_mc}, for a system with $20 \times 20$ unit cells, equivalent to $800$ spins. The spin-lattice coupling was chosen as $g = 0.5$ meV. The error bars show the statistical uncertainty in the average magnetization.}
    \label{fig:lambda_t}
\end{figure}

\begin{figure}
    \centering
    \begin{tikzpicture}
     \node at ( 0, 2.5) {\includegraphics[width=\columnwidth]{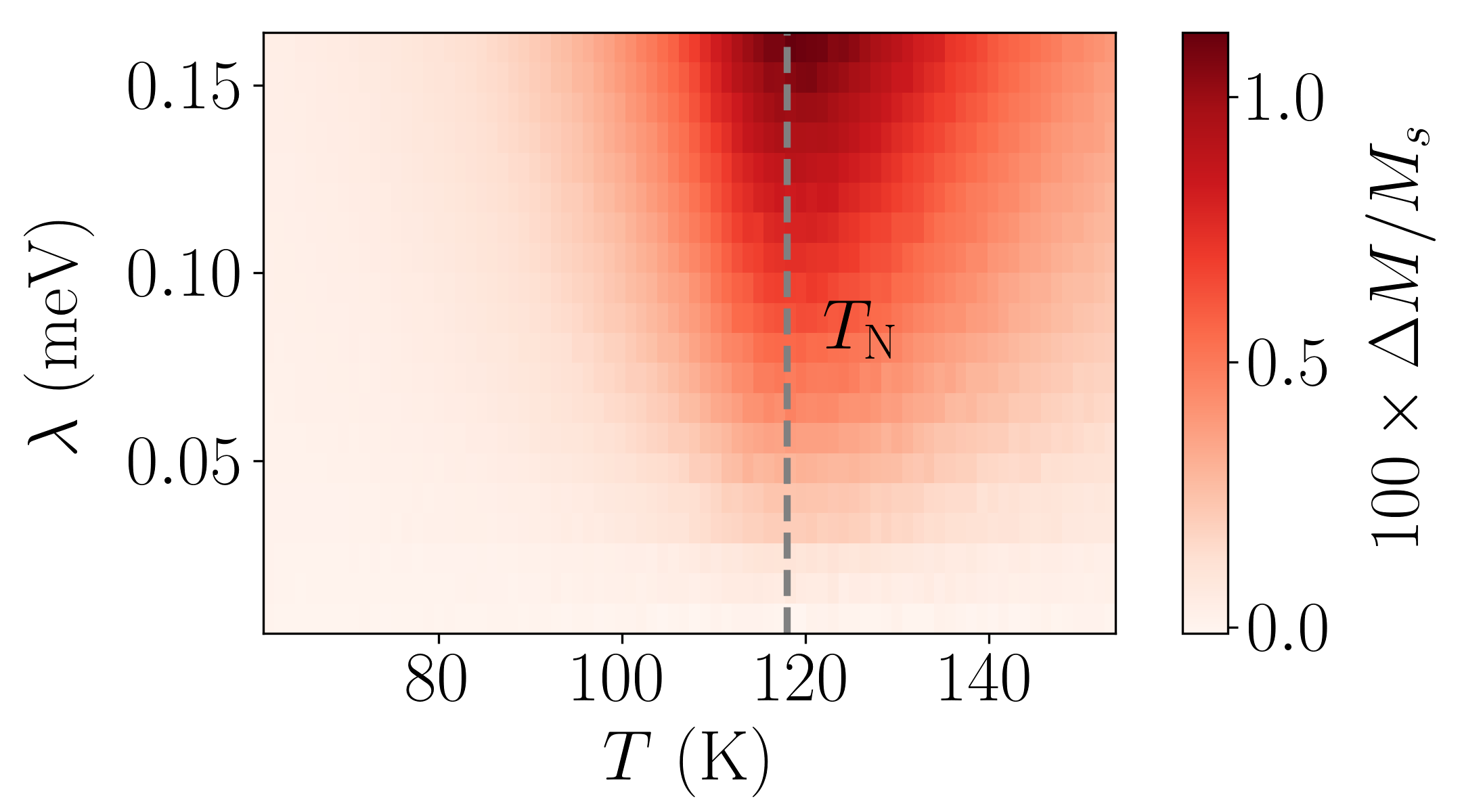}};
     \node at ( 0,-2.5) {\includegraphics[width=\columnwidth]{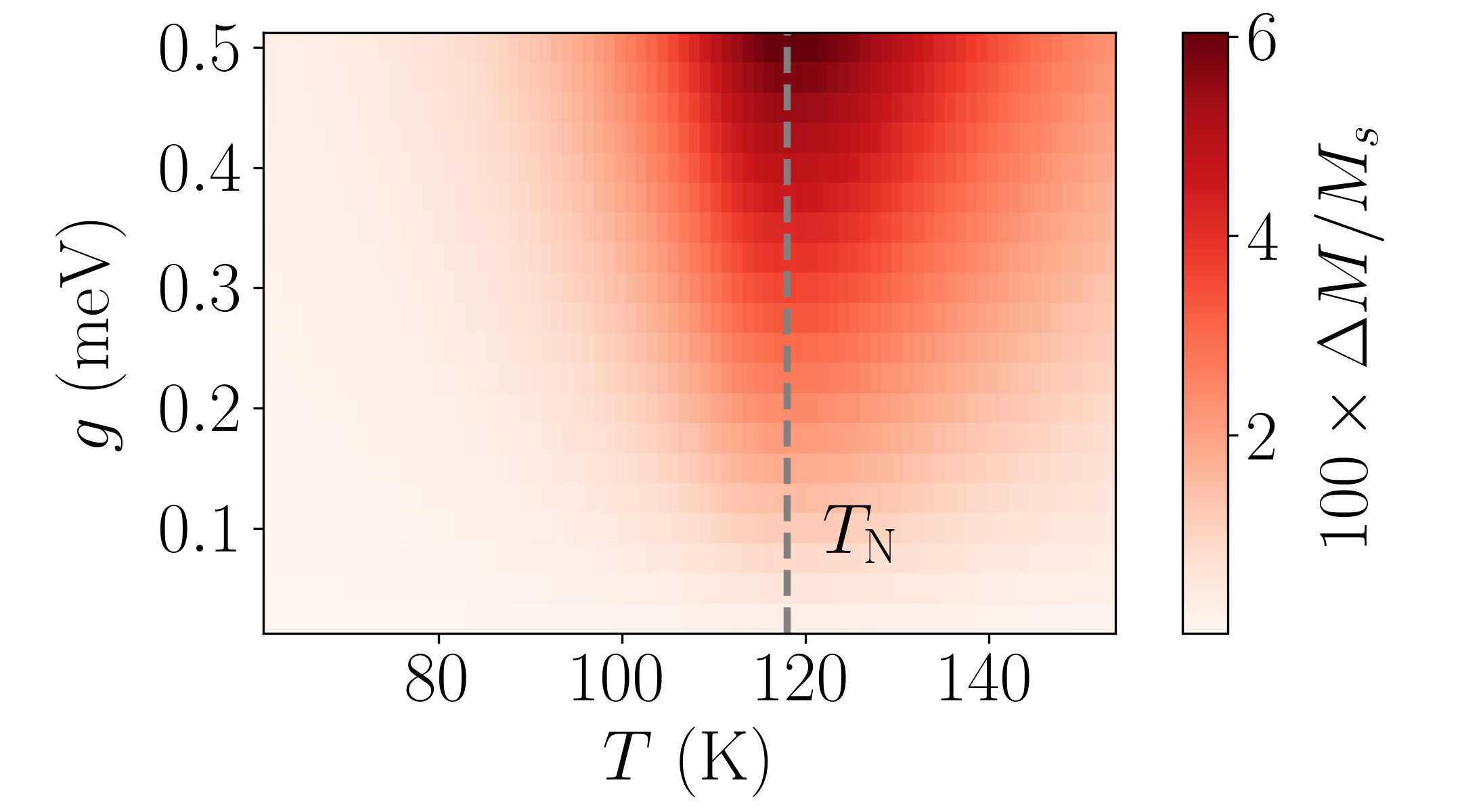}};

     \node at (-2.35, 4.2) {\large\bf a};
     \node at (-2.35,-0.8) {\large\bf b};
    \end{tikzpicture}
    \caption{Dependence of the induced magnetization on ({\bf a}) the phonon-cavity coupling $\lambda$, assuming a spin-phonon coupling $g = 0.5$ meV, and on ({\bf b}) the spin-phonon coupling $g$, assuming a phonon-cavity coupling $\lambda = 0.5$ meV. The results were obtained via the self-consistent Monte Carlo scheme described in Supplementary Section~\ref{app:sc_mc}, for a system with $20 \times 20$ unit cells, equivalent to $800$ spins.}
    \label{fig:magnetization_coupling}
\end{figure}

Using the self-consistent Monte Carlo scheme developed in the previous section, we calculate the dependence of the induced magnetization on the Raman-cavity coupling $\lambda$. The results are shown in Fig.~\ref{fig:lambda_t} and Fig.~\ref{fig:magnetization_coupling}, and were obtained for a spin system with $20 \times 20$ unit cells at a spin-lattice coupling of $g = 0.5$ meV and for temperatures around the N\'eel temperature $T = T_{\rm N}$. The linear dependence of $M$ on $\lambda$ is consistent with the analytical results of the main text, obtained from a mean-field treatment of Ginzburg-Landau theory.

In Fig.~\ref{fig:magnetization_coupling} we also show the dependence of $M$ on the spin-phonon coupling $g$, at a fixed value of $\lambda = 0.5$ meV.


\section{Finite size scaling}\label{app:scaling}
We here discuss the behavior of the magnetic order parameters and susceptibilities in a finite simulation volume $N = L^2$. In the thermodynamic limit $N \to \infty$, it is known that for $T \approx T_c$ the dominant order parameter and susceptibility satisfy
\begin{align}
 L(\tau) &\sim |\tau|^\beta \quad \text{for} \quad \tau < 0 \\
 \chi_L(\tau) &\sim |\tau|^{-\gamma},
\end{align}
where $\tau = (T - T_c)/T_c$ is reduced temperature. Alternatively, we can write these quantities in terms of the correlation length $\xi \sim |\tau|^{-\nu}$ as,
\begin{align}
 L(\tau) &\sim \xi^{-\beta/\nu} \phi_L(\xi/L) \quad \text{for} \quad \tau < 0 \\
 \chi_L(\tau) &\sim \xi^{\gamma/\nu} \phi_\chi(\xi/L).
\end{align}
where $\phi_i(x)$ are scaling functions such that $\phi(x) \to \text{const.}$ as $x \to 0$ and $\phi(x) \to x^{-\gamma/\nu}$ as $x \to \infty$. Defining a new scaling function through $\phi(x) = x^{-\gamma/\nu} \varphi(x^{-1/\nu})$, such that the dependence on the finite length scale $L$ becomes explicit, we find
\begin{align}
 L(\tau) &\sim L^{-\beta/\nu} \varphi_L(L^{1/\nu} \tau) \quad \text{for} \quad \tau < 0 \\
 \chi_L(\tau) &\sim L^{\gamma/\nu} \varphi_\chi(L^{1/\nu} \tau).
\end{align}
Here $\varphi(x) \to \text{const.}$ as $x \to 0$ and $\varphi(x) \to L^{-\gamma/\nu} |\tau|^{-\gamma}$ as $x \to \infty$.

In a truly finite system (i.e. a system finite along all dimensions), the critical divergence of the order parameter is cut off when $\xi \sim L$, leading to a rounding of the divergences. Phrased in terms of the reduced temperature, the singularities will be rounded over a region of $\tau$ such that $\xi(\tau) \gtrsim L$. In such a system, $\varphi(x)$ will be a smooth function with a peak at some location $x = x_0$ and with a characteristic width $\Delta x$. Therefore, $\chi$ will have a peak at the location $T = T_c + x_0 L^{-1/\nu}$ with a magnitude $L^{\gamma/\nu}$ and a width $\Delta x L^{-1/\nu}$. Around the critical temperature, we can expand the susceptibility as
\begin{align}
 \chi_L(\tau) &\sim L^{\gamma/\nu} (1 + a |\tau| + b |\tau|^2),
\end{align}
with some coefficients $a$ and $b$. We note that the exponent $\gamma/\nu = 2 - \eta$, with $\eta$ the anomalous dimension, such that $\chi_l/V \sim L^{2-\eta}/V$.

As discussed in the main text, a theory that combines a mean-field treatment of the Ginzburg-Landau equations with renormalization group values of the critical exponents, predicts a susceptibility density $\chi_l/V$ that vanishes in the thermodynamic limit for $d \geq 2$. Therefore, such a theory predicts an induced magnetization that vanishes in the thermodynamic limit. This is in contrast to our Monte Carlo simulations, which predict a magnetization independent of system size (see Fig.~\ref{fig:mc_data}). Since the Monte Carlo simulations include higher order terms in the Ginzburg-Landau theory, we interpret the discrepancy as being due to these terms. In particular, we note that using the mean-field exponents for $\chi_L$ gives an anomalous dimension $\eta = 0$, such that the magnetization becomes independent of volume close to $T_{\rm N}$ in line with the Monte Carlo data. We conjecture that this results also holds in the full Ginzburg-Landau theory, as supported by our Monte Carlo data.


\section{First principles calculations}\label{app:first_principles}
To obtain the phonon-SPP coupling of FePS$_3$, we performed first principles simulations with the {\sc abinit} electronic structure code~\cite{Gonze2020TheDevelopments,Gonze1997First-principlesAlgorithm,Amadon2008Plane-waveOrbitals,Torrent2008ImplementationPressure}. We used the local density approximation with projector augmented wave (PAW) pseudopotentials, a plane wave cut-off of $20$ Ha and $40$ Ha respectively for the plane wave and PAW part, and included an empirical Hubbard $U$ of $2.7$ eV on the Fe $d$-orbitals as self-consistently determined in the {\sc octopus} electronic structure code via the ACBN0 functional. A $\Gamma$-centered Monkhorst-Pack grid with dimensions $8\times 6 \times 8$ was used to sample the Brillouin zone. The ground state was found to have zig-zag antiferromagnetic order with spins aligned along the $z$-axis. Phonon frequencies and eigenvectors were calculated with {\sc abinit} assuming a ferromagnetic interlayer coupling, after relaxing the atomic positions and stresses to below $10^{-6}$ Ha/Bohr. The Raman tensor was obtained from the change in the dielectric tensor with respect to small displacements of the $3.27$ THz phonon mode, calculated using density functional perturbation theory. From these calculations we find $\partial\boldsymbol\epsilon/\partial Q_i = 0.33 a_0^{-1}$ with $a_0$ the Bohr radius.


\end{document}